\documentclass[a4paper,prd,11pt,noshowkeys,notitlepage,nofootinbib,superscriptaddress]{revtex4}
\pdfoutput=1  
\usepackage[utf8]{inputenc}
\usepackage{ae,aecompl}
\usepackage{amsmath,amssymb,mathrsfs}
\usepackage{graphicx}
\usepackage{ccnishi}
\usepackage{dsfont}
\usepackage{slashed}
\usepackage{units}

\usepackage[usenames,dvipsnames]{xcolor} 
\usepackage{hyperref}
\hypersetup{
    colorlinks=true,        
    linkcolor=purple,       
    citecolor=JungleGreen,  
    filecolor=magenta,      
}

\providecommand{\cM}{\mathscr{M}}

\providecommand{\cY}{\mathscr{Y}}
\providecommand{\cO}{\mathcal{O}}

\providecommand{\btheta}{\bar{\theta}}

\providecommand{\ckmsm}{V_{\rm ckm}^{\rm sm}}

\providecommand{\hY}{\hat{Y}}

\providecommand{\lag}{\mathscr{L}}

\providecommand{\eps}{\epsilon}

\makeatletter
\newcommand*\bigcdot{\mathpalette\bigcdot@{.5}}
\newcommand*\bigcdot@[2]{\mathbin{\vcenter{\hbox{\scalebox{#2}{$\m@th#1\bullet$}}}}}
\makeatother

\begin{document}
\title{
Consequences of vector-like quarks of Nelson-Barr type
}
\author{A.~L.~Cherchiglia}
\email{adriano.cherchiglia@ufabc.edu.br}
\affiliation{Centro de Ci\^{e}ncias Naturais e Humanas,
Universidade Federal do ABC, Santo Andr\'{e}-SP, Brasil}
\author{C.~C.~Nishi}
\email{celso.nishi@ufabc.edu.br}
\affiliation{Centro de Matem\'{a}tica, Computa\c{c}\~{a}o e Cogni\c{c}\~{a}o,
Universidade Federal do ABC, Santo Andr\'{e}-SP, Brasil}

\begin{abstract}
The Nelson-Barr mechanism to solve the strong CP problem requires vector-like quarks (VLQs) to transmit the spontaneous CP breaking to the SM.
We study the scenario where only these VLQs are within reach at the TeV scale while the spontaneous CP breaking sector is inaccessible.
We investigate how these VLQs of Nelson-Barr type differ from generic VLQs and find from parameter counting that one less parameter is needed.
In particular, for one VLQ of Nelson-Barr type, there is only one CP odd quantity that is responsible for all CP violation.
In this case, we solve the technical problem of parametrizing only the new physics parameters while keeping the SM parameters as independent inputs.
For one down-type VLQ, the model is largely flavor safe because the VLQ couplings to the SM up quarks and the $W$ are hierarchically smaller for lighter quarks.
\end{abstract}
\maketitle
\section{Introduction}
\label{sec:intro}

The strong CP problem is a naturalness problem coming from the clash between a CP violating parameter $\btheta$ that is experimentally constrained to be very small ($\sim 10^{-10}$) and the expectation of an order one parameter coming from two sources: the unknown contribution of the nontrivial vacuum structure of QCD and the well measured order one CP phase of the quark sector residing in the CKM matrix (see \cite{Kim:2008hd} for a review).

The explanation for the strong CP problem usually invokes two possibilities: (i) 
the promotion of $\btheta$ to a dynamical field -- the axion -- which couples to the QCD gluon potential and then is dynamically driven to zero in the potential minimum\,\cite{PQ};
(ii) CP (or P) is indeed a fundamental symmetry and its violation manifests itself only through spontaneous breaking at lower energies making $\btheta$ calculable and to arise only at loop level, potentially justifying its tiny value\,\cite{scpv:others,nelson,barr}.\,\footnote{%
A third solution of a massless up quark\,\cite{massless.u} is dismissed because it is strongly disfavoured by lattice calculations\,\cite{lattice}.
However, see e.g. \cite{bardeen} and \cite{mu=0:high}.
}

Undoubtedly, the first solution (i) is the most popular\,\cite{nardi}. In recent times, the interest on axions is being revived in connection to the dark matter problem\,\cite{essig}, to the hierarchy problem (the relaxion)\,\cite{relaxion}, to flavor physics (flavored axions)\,\cite{king.flavor.axion,astrophobic} and to solve a collection of problems in one stroke\,\cite{smash}.
Axion-like particles, that are unrelated to the strong CP problem, retain many phenomenological similarities to axions and may also solve many of the mentioned problems\,\cite{accidental.axion}.

We will focus instead on the less popular approach (ii), although connections to other problems such as the hierarchy problem\,\cite{NB-relaxion} are being explored.
Within approach (ii), the Nelson-Barr mechanism is one of the simplest ways to guarantee $\btheta=0$ at tree-level from explicit CP conservation\,\cite{nelson,barr}.%
\footnote{See other proposals in Refs.\,\cite{scpv:others:recent,cp-texture,meade:SFV}.}
This mechanism requires a spontaneously CP violating scalar sector at some $\Lambda_{CP}$ scale and vector-like quarks (VLQs) at much lower scales which transmit CP violation to the SM.
The big challenge is to emulate the relatively large explicit CP violation of the SM\footnote{CP violation in the SM is quantitatively small due to the small mixing angles.}
from spontaneous CP violation and yet make the loop contributions to $\btheta$ small.
See Ref.\,\cite{dine} for the naturalness issues involved.
As a related idea, if spontaneous CP breaking is linked to flavor violation, FCNC may be naturally suppresed to allow new physics at the TeV scale with flavor nonuniversal and nonhierarchical features\,\cite{meade:SFV}. Our scenario differs from this case.

If the Nelson-Barr mechanism is at work in Nature,
one can envision a scenario where only these VLQs are within reach, at the TeV scale, while the scalar sector lies much higher, inaccessible to our probes.
The questions that follow are how to define this scenario, how to constrain it and how to distinguish these VLQs from generic VLQs unrelated to the strong CP problem or the origin of CP violation in the SM.
These are the questions that motivate us here.
To avoid ambiguity, we denote these VLQs arising from the Nelson-Barr mechanism as VLQs of Nelson-Barr type or NB-VLQs.
Differently from generic VLQs, the expectation that VLQs of Nelson-Barr type should leave some trace at low energy relies on the characteristic of this scenario that these VLQs cannot decouple completely from the SM because they \emph{must} transmit the CP violation to the SM.

The minimal case of one NB-VLQ of down type coincides with the minimal implementation of Bento, Branco and Parada\,\cite{BBP} after spontaneous CP breaking takes place.
After studying the generic case of NB-VLQs of down type,
we study this case in some detail and seek a parametrization that would reproduce the SM flavor structure for quarks, including its CP violation.
This parametrization and the identification of the number of CP odd quantities are the main novel theoretical results we bring. With an appropriate parametrization in hand, some phenomenological aspects are analyzed.

This paper is organized as follows: in Sec.\,\ref{sec:NB} we define what we mean by vector-like quarks of Nelson-Barr type and show that one less parameter is needed to describe them compared to generic VLQs.
This definition is connected with the usual notation of VLQs in Sec.\,\ref{sec:diag.1}.
We review the couplings of these heavy quarks to the SM gauge bosons $W$ and $Z$ in 
Sec.\,\ref{sec:obs}.
Section \ref{sec:CPV} analyzes the intriguing question of which parameters are responsible for CP violation.
In Sec.\,\ref{sec:irred.flavor} we show that NB-VLQs cannot decouple completely and an irreducible amount of flavor changing neutral coupling to $Z$ inevitably remains.
For one NB-VLQ, we solve in Sec.\,\ref{sec:param} the technical problem of how to parametrize this model incorporating the flavor structure and CP violation of the SM quark sector. This parametrization is denoted as the seesaw parametrization.
The phenomenology of the model is analyzed in Sec.\,\ref{sec:pheno} focusing on flavor constraints for the case of one NB-VLQ.
The conclusions are presented in Sec.\,\ref{sec:conclu} and the appendices contain auxiliary material.

\section{Vector-like quarks of Nelson-Barr type}
\label{sec:NB}

To define what we mean by VLQs of Nelson-Barr type (NB-VLQs), we start by describing generic VLQs. We consider only the case of singlets of $SU(2)_L$ of charge $-1/3$ denoted by $B_{rL},B_{rR}$, $r=1,\dots,n_B$.
The case of singlet quarks of charge $2/3$ or the case of doublets can be equally considered.
Note that in order to transmit the CP violation to the SM at the renormalizable level the VLQs need to be singlets, doublets or triplets and there are only 8 possibilities.\footnote{Non-renormalizable interactions with VLQs have also been considered\,\cite{non.ren.VLQ} which enlarges the type of multiplets allowed.}
See e.g.\ Refs.\,\cite{delaguila:eft,ligeti.wise} for the complete quantum numbers.

We can write the SM Yukawa Lagrangian with the $N_f=3$ chiral quark families as
\eq{
\label{yuk:SM}
-\lag\supset\bar{q}_{iL}Y^d_{ij} Hd_{jR}+\bar{q}_{iL}Y^u_{ij} \tilde{H}u_{jR}+h.c.
}
In explicit situations, we will use the basis where $Y^u$ is diagonal and denote it as $\hY^u$.
The VLQs then couple as
\eq{
\label{yuk:VLQ}
-\lag \supset \bar{q}_{iL} Y^B_{ir}H\,B_{rR} 
+ \bar{B}_{rL}M^B_{rs} B_{sR}+h.c.,
}
where $M^B$ is expected to be much larger than the electroweak scale.
We already eliminated some terms by rotating in the space $(d_R,B_R)$.
The total number of physical parameters contained in $Y^u,Y^d,Y^B,M^B$ is\,\footnote{%
Take, for instance, the basis where $M^B$ and $Y^u$
are diagonal.}
\eq{
\label{N.para:VLQ}
N_{\rm param}=N_f^2+1+2N_fn_B\,,
}
which amounts to 6 additional parameters for each VLQ compared to the SM $N_f^2+1=10$ parameters, corresponding to the usual 6 masses, 3 mixing angles and one CP phase.
If we are restricted to only one type of VLQ, the parameter counting is also the same for up-type singlets.
By leaving the number of VLQs free we can study subsectors of models in which multiple heavy quarks are required. For example, the implementation of the minimal flavor violation with the presence of VLQs requires at least three of them\,\cite{VLQ:MFV}.

Now we will define VLQs to be of Nelson-Barr type
when the theory defined by \eqref{yuk:SM} and \eqref{yuk:VLQ} arises from a structure of the form
\eqali{
\label{yuk:NB}
-\lag&=\bar{q}_{iL}\cY^d_{ij} Hd_{jR}+\bar{q}_{iL}\cY^u_{ij} \tilde{H}u_{jR}
\cr
&\quad +\ 
\bar{B}_{rL}\cM^{Bd}_{rj} d_{jR}+\bar{B}_{rL}\cM^B_{rs} B_{sR}+h.c.,
}
with the additional requirement that $\cY^u,\cY^d,\cM^{B}$ are \textit{real} matrices and only $\cM^{Bd}$ is complex.%
\footnote{%
We could base this definition on the imposition of CP symmetry and a $\ZZ_2$ symmetry where only $B_R,B_L$ are odd so that CP and $\ZZ_2$ may only be broken softly and simultaneously as in the Bento-Branco-Parada (BBP) model\,\cite{BBP}.
This means CP breaking is triggered by $\ZZ_2$ odd scalars.
For exampe, the term $\bar{q}_LHB_R$ is not allowed because $\ZZ_2$ breaking is not soft.
Other abelian $U(1)$ or $\ZZ_n$ symmetries\,\cite{dine} could be used but a $\ZZ_2$ is always definable.
}

In Nelson-Barr type models CP is a fundamental symmetry (hence real parameters) which is only spontaneously broken (hence complex $\cM^{Bd}$) at a scale much higher than the VLQ masses\,\cite{nelson,barr}.
Considering that the effective mass matrix arising from \eqref{yuk:NB} has real determinant, the strong CP parameter $\btheta$ vanishes at tree level and the problem is solved if the loop corrections are small enough.
In our definition, the complex $\cM^{Bd}$ parametrizes our ignorance about the sector responsible for spontaneous CP violation and being the coefficient of a dimension three term it indicates that CP is only softly broken.\footnote{%
A similar but more specific scenario was proposed in Ref.\,\cite{lavoura}.%
}
Thus the VLQs are responsible for transmitting the CP violation to the SM.
For the CP breaking sector, one singlet complex scalar can do the job and the $n_B=1$ case was proposed by Bento, Branco and Parada (BBP)\,\cite{BBP}.
For $n_B=2$, one can even make the one-loop contribution to $\btheta$ vanish by imposing a non-conventional CP\,\cite{NB:CP4}.

We can see NB-VLQs comprise only a subclass of general VLQs by counting the number of physical parameters in the Lagrangian \eqref{yuk:NB}:
\eq{
\label{N.param:NB}
N_{\rm param}\big|_{\rm NB}=\ums{2}N_f(N_f+3)+2N_fn_B\,.
}
Here, instead of unitary rotations, we must use real rotations to maintain the structure of \eqref{yuk:NB}.
The number for $n_B=0$ amounts to $N_{\rm param}\big|_{\rm NB}=9$ which corresponds to the 6 masses and 3 mixing angles of the unrealistic CP conserving SM.
The number of additional parameters for each VLQ is still 6.
So a model of NB-VLQs has \emph{one less} parameter than a generic model of VLQs.
For $n_B=1$, of the 15 parameters, 10 should account for the SM ones. The additional five parameters describe physics beyond the SM.
One of our main goals is to seek ways to distinguish generic VLQs from NB-VLQs.

%

\section{Partial diagonalization}
\label{sec:diag.1}

To specify how NB-VLQs differ from generic VLQs, we need to change basis from \eqref{yuk:NB} to \eqref{yuk:VLQ}. 
This is achieved by the rotation in the $3+n_B$ dimensional space
\eq{
\mtrx{d_R\cr B_R}\to W_R\mtrx{d_R\cr B_R}\,,
}
restricted to 
\eq{
\det W_R=1\,,
}
to avoid transferring complex phases to $\theta$ of QCD through chiral rotations.
The unitary matrix $W_R$ is defined by
\eq{
\label{WR:def}
\left(\begin{array}{c|c}
\cM^{Bd} &\cM^B
\end{array}\right)
W_R=
\left(\begin{array}{c|c}
0_{n_B\times 3} &M^B
\end{array}\right)
\,,
}
and hence it is generically complex due to $\cM^{Bd}$.
We can write $W_R$ as a collection of column vectors
\eq{
W_R=\left(\begin{array}{c|c|c|c}
u_1 & u_2 &\cdots & u_{3+n_B}
\end{array}\right)\,,
}
where the last $n_B$ vectors $u_{3+a}$, $a=1,\dots,n_B$, of size $(3+n_B)$, 
because of unitarity of $W_R$,
form an orthonormal basis for the space spanned by the $n_B$ columns of\,\footnote{%
This can be achieved by, e.g., orthogonalization.
}
\eq{
\label{nB:columns}
\left(\begin{array}{c}
{\cM^{Bd}}^\dag \cr {\cM^B}^\tp
\end{array}\right)
\,.
}
The three vectors $u_1,u_2,u_3$, should be chosen to be orthogonal to $u_{3+a}$, $a=1,\dots,n_B$.

Let us define sub-blocks of $W_R$ as 
\eq{
\label{WR}
W_R=
\left(\begin{array}{c|c}
W_R^{dd} & W_R^{dB}
\cr\hline
W_R^{Bd} & W_R^{BB}
\end{array}\right)\,,
}
where $W_R^{dd}\sim 3\times 3$ and $W_R^{dB}\sim 3\times n_B$ are the parts relevant for the coupling to the SM while $W_R^{BB}\sim n_B\times n_B$ and $W_R^{Bd}\sim n_B\times 3$ will be only implicit in the heavy VLQ mass matrix.
Obviously, being sub-matrices of a unitary matrix, they 
do not have to be unitary themselves ($W_R^{dB}$ may not even be square).

We can finally write the NB-VLQ Lagrangian \eqref{yuk:NB} in the form \eqref{yuk:VLQ} with the identification
\eq{
\label{Yd:YB}
Y^d=\cY^d W_R^{dd}\,,\quad
Y^B=\cY^d W_R^{dB}\,,
}
together with the relation \eqref{WR:def} which defines $M^B$.
So the Yukawa couplings for SM quarks and the VLQ portal coupling to the SM are not independent.
In fact, unitarity of $W_R$ implies the sum rule
\eq{
\label{sum.rule}
Y^d{Y^d}^\dag+Y^B{Y^B}^\dag=\cY^d{\cY^d}^\tp\sim \text{$3\times 3$ real}.
}

Explicit expressions for the sub-blocks of $W_R$ in \eqref{WR} would be helpful for a general analysis.
The zero block in \eqref{WR:def} allows to relate
\eq{
\label{W:Bd}
W_R^{Bd}=-\cM^{B^{\mss{-1}}}\cM^{Bd}\,W^{dd}_R\,.
}
If we use the VLQ mass matrix $M^B$ as input, we can explicitly write
\eq{
\label{W:db;BB}
W_R^{dB}={\cM^{Bd}}^\dag \big({M^B}^\dag\big)^{-1}\,,
\quad
W_R^{BB}={\cM^{B}}^\dag \big({M^B}^\dag\big)^{-1}\,,
}
where $M^B$ should obey
\eq{
\label{HB}
M^B M^{B\dag}=
H_B\equiv \cM^{Bd}{\cM^{Bd}}^\dag + \cM^{B}{\cM^{B}}^\tp\,.
}
It is easy to check eq.\,\eqref{WR:def} is satisfied
and unitarity for the last $n_B$ columns of \eqref{WR} can be checked as well.
Hence, the relations \eqref{W:db;BB} for the sub-blocks are uniquely defined because they compose an orthonormal basis for the space \eqref{nB:columns}; the only freedom is a unitary change of basis which can be left implicit in the definition of $M^B$.
We could further write $M^B=H_B^{1/2}$ if we choose a hermitian $M^B$.
Since $M^B$ is the mass matrix for the VLQs, neither $M^B$ nor $H_B$ can be singular and their inverses are always meaningful.

The relations \eqref{W:Bd} and \eqref{W:db;BB} leave only the sub-block $W_R^{dd}$ implicitly defined in $W_R$.
But $W_R^{dd}$ is only defined modulo further unitary rotations from the right since the orthonormality of the upper row of blocks in \eqref{WR} implies
\eq{
\label{W:dd:1}
W_R^{dd}{W_R^{dd}}^\dag=\id_3-{\cM^{Bd}}^\dag H_B^{-1}\cM^{Bd}\,.
}
See appendix \ref{ap:partial} for more relations.
Note that $W_R^{dd}$ cannot be singular because it is part of the SM Yukawa.

Now the SM Yukawa in \eqref{Yd:YB} is 
\eq{
\label{Yd:2}
Y^d{Y^d}^\dag=\cY^d\big(\id_3-{\cM^{Bd}}^\dag H_B^{-1}\cM^{Bd}\big){\cY^d}^\tp
\,,
}
which is the leading expression obtained from the quark seesaw in BBP type models\,\cite{BBP,NB:CP4}.
Notice that the right-hand side is explicitly written in terms of parameters of the NB Lagrangian \eqref{yuk:NB} and the complexity of $\cM^{Bd}$ is what generates the CKM phase in $V_{d_L}$ which diagonalizes 
\eq{
\label{ckm:sm}
{V^\dag_{d_L}} Y^d{Y^d}^\dag V_{d_L}=\frac{2}{v^2}\diag(m_d^2,m_s^2,m_b^2)\,.
}
So $V_{d_L}\approx \ckmsm$, i.e., the CKM matrix of the SM.

Using the previous relation, we can see that the sum rule \eqref{sum.rule} strongly relates the VLQ coupling $Y^B$ with the SM coupling $Y^d$.
One example is the relation
\eq{
\im\{Y^B{Y^B}^\dag\}=-\im\{Y^d{Y^d}^\dag\}=
-\im\{V_{d_L}(\hY^d)^2{V^\dag_{d_L}}\}
\,
}
which is completely fixed for $V_{d_L}\approx \ckmsm$ while
$(\hY^d)^2$ is the right-hand side of \eqref{ckm:sm}.

\section{Observables: Couplings to $W,Z$ and higgs}
\label{sec:obs}

Below the electroweak scale, the presence of VLQs can only be inferred from their couplings with the gauge bosons $W,Z$ and the higgs.
In particular,
the presence of flavor changing neutral currents coupled to $Z$ is a well known consequence\,\cite{VLQ:FCNC}.
In the weak eigenstate basis, where the Yukawa Lagrangian in Eqs.\,\eqref{yuk:SM} and \eqref{yuk:VLQ} are valid, these couplings are given by
\eqali{
-\lag_{W}&=\frac{g}{\sqrt{2}}\bar{u}_{iL}\gamma^{\mu}d_{iL}W_{\mu}^{+}+h.c.
\cr
-\lag_{Z}&=\frac{g}{2c_W}\Big(\bar{u}_{iL}\gamma^{\mu} u_{iL}
-\bar{d}_{iL}\gamma^{\mu} d_{iL}-2s^2_W J^\mu_{e.m.}\Big)Z_\mu
}
where the implicit sum over $i,j=1,2,3$ are for doublet quarks $q_{iL}=(u_{iL},d_{iL})^\tp$ and $J^\mu_{e.m.}$ is the electromagnetic current containing 
usual quarks and VLQs.
The coupling with the higgs can be read off from the Yukawa Lagrangian.

The mass eigenstate basis is reached with the transformations
\eq{
\label{diag:VLQ}
\mtrx{d_{L} \cr B_{L}} \to 
U_{d_L}\mtrx{d_{L} \cr B_{L}}
\,,\quad
\mtrx{d_{R} \cr B_{R}} \to 
U_{d_R}\mtrx{d_{R} \cr B_{R}}\,,
}
that diagonalizes\,%
\footnote{To be precise, $\hat{M}^B$ in the right-hand side is not the diagonalized form of $M^B$ in the left-hand side; they should be attributed different symbols. But they coincide within the seesaw approximation.}
\eq{
\label{diag.2}
U_{d_L}^\dag
\mtrx{\frac{v}{\sqrt{2}}Y^d & \frac{v}{\sqrt{2}}Y^B
\cr 0 & M^B}
U_{d_R}
=\mtrx{\hat{M}^d & \cr & \hat{M}^B}\,,
}
where $\hat{M}^d=\diag(m_d,m_s,m_b)$ and $\hat{M}^B=\diag(M^B_1,\dots,M^B_{n_B})$.
We consider $Y^u=\hY^u$ to be already diagonal.
Since we use the parameters in the Lagrangian in Eqs.\,\eqref{yuk:SM} and \eqref{yuk:VLQ}, we are treating generic VLQs at this point. For NB-VLQs some of these parameters are correlated and we only treat this subclass in the end of the section.

This diagonalization can be performed approximately using a seesaw expansion:
\eq{
\label{seesaw}
U_{d_L}\approx \mtrx{\id_3-\ums{2}\theta_L\theta_L^\dag & \theta_L \cr -\theta_L^\dag & \id_{n_B}-\ums{2}\theta_L^\dag\theta_L}
\mtrx{V_{d_L} & \cr & V_{B_L}}
\,,
}
where
\eq{
\label{thetaL}
\theta_L=\frac{v}{\sqrt{2}}Y^B{M^B}^{-1}\,,
}
and $V_{d_L}$, $V_{B_L}$ are unitary matrices. Notice that $\theta_L\ll 1$ when $M^B\gg v$ and it can be used as an expansion parameter. A systematic expansion can be performed order by order; see example for Majorana neutrinos\,\cite{seesaw:exp}.
The $3\times 3$ block of the right-hand side of \eqref{diag.2} containing the SM quark masses is simply
\eq{
\label{seesaw.block.1}
\big(\hat{M}^d\big)^2=V_{d_L}^\dag\left(\frac{v^2}{2}Y^d{Y^d}^\dag\right)V_{d_L}\,.
}
Note that the contribution from $Y^B$ is canceled within this approximation.

From eq.\,\eqref{seesaw.block.1}, it is clear that the matrix $V_{d_L}$ diagonalizes $Y^d{Y^d}^\dag$.
Likewise, $V_{B_L}$ diagonalizes $M^B{M^B}^\dag$.
So $V_{d_L}$ is approximately equal to $\ckmsm$ when $Y^u$ is diagonal.
For $U_{d_R}$ we can use the same expansion \eqref{seesaw} with $\theta_L$ replaced by
\eq{
\label{thetaR}
\theta_R
=\frac{v}{\sqrt{2}}{Y^d}^\dag\theta_L{{M^B}^{\dag}}^{-1}\,.
}
The matrices $V_{d_L},V_{B_L}$ are also replaced by $V_{d_R},V_{B_R}$ and their definitions should be adapted accordingly.
We can see that if $\theta_L$ is roughly of order $\eps$, then $\theta_R$ is of order $\eps^2$ for order one yukawas\,\cite{saavedra:handbook}.

The couplings in the mass eigenstate basis are\,\cite{saavedra:flavor}
\eqali{
\label{L:W}
-\lag_{W}&=
\frac{g}{\sqrt{2}}\bar{u}_{iL}\gamma^{\mu}\big(V_{ij}d_{jL}+V_{i,3+a}B_{aL}\big)
W_{\mu}^{+}+h.c.,
\cr
-\lag_{Z}&=\frac{g}{2c_W}\left[\bar{u}_{iL}\gamma^{\mu}u_{iL}-\bar{d}_{iL}\gamma^{\mu}X^{d}_{ij}d_{jL}
-\bar{B}_{aL}\gamma^{\mu}X^{d}_{3+a,3+b}B_{bL}
\right.
\cr&\hspace{3em}
\left.
-2s_W^{2}J_{e.m.}^{\mu}
-\big(\bar{d}_{iL}\gamma^{\mu}X^{d}_{i,3+a}B_{aL}+h.c.\big)
\right]Z_{\mu}\,,
}
where $a$ runs through $1,\dots,n_B$.
The rectangular matrix $V\sim 3\times(3+n_B)$ which describes the quark couplings to $W$ is given by
\eq{
\label{V}
V=U_{u_L}^\dag P U_{d_L}\approx
V_{d_L}
\left(\begin{array}{c|c}
\id_3-\ums{2}\delta X^d & \Theta
\end{array}\right)\,,
}
where $P$ is a $3\times (3+n_B)$ projection matrix, nonvanishing only for $P_{11}=P_{22}=P_{33}=1$, and
\eq{
\label{Theta}
\Theta\equiv V_{d_L}^\dag \theta_L V_{B_L}\,,
\quad
\delta X^d\equiv \Theta\Theta^\dag\,.
}
The square matrix $X^d$ of size $(3+n_B)$ that describes the FCNC coupling to $Z$ is
\eq{
\label{X^d}
X^d=V^\dag V
\approx 
\left(\begin{array}{c|c}
\id_3-\delta X^d & \Theta
\cr\hline
*  & \Theta^\dag\Theta
\end{array}\right)\,.
}
So we see that all couplings of VLQs to gauge bosons depend solely on the matrix $\Theta$ within the seesaw approximation.

The higgs coupling in the mass eigenstate basis is
\eq{
-\lag_h=\frac{h}{v}\mtrx{\bar{d}_L&\bar{B}_L}N^d\mtrx{d_R\cr B_R}+h.c.\,,
}
where, at leading order, 
\eq{
N^d\approx
\mtrx{\big(\id_3-\delta X^d\big)\hat{M}^d & 
	\Theta\hat{M}^B
	\cr
	\Theta^\dag \hat{M}^d
	& 	\Theta^\dag\Theta\hat{M}^B
}\,.
}
The first term in the upper-left block is the standard coupling proportional to the quark masses.
The term in the upper-right block is the dominant higgs coupling to VLQs and induces the decay $B_R\to d_L+h$.

At high energy the equivalence theorem tell us that the decay to longitudinal gauge bosons $Z_L~(\varphi^0)$ and $W_L~(\varphi^+)$ are induced by the couplings
\eq{
\bar{d}_{iL}\bigg(\Theta \frac{\sqrt{2}}{v}\hat{M}^B\bigg)_{ia}B_{aR}\frac{(h+i\varphi^0)}{\sqrt{2}}
+
\bar{u}_{iL}\bigg(\ckmsm \Theta \frac{\sqrt{2}}{v}\hat{M}^B\bigg)_{ia}B_{aR}\varphi^+
\,.
}

Now we can specialize to NB-VLQs. 
Firstly, the $4\times 4$ mass matrix in \eqref{diag.2},
\eq{
\mtrx{\frac{v}{\sqrt{2}}Y^d & \frac{v}{\sqrt{2}}Y^B \cr 0 & M^B },
}
comes from 
\eq{
\mtrx{\frac{v}{\sqrt{2}}\cY^d & 0 \cr\cM^{Bd} & \cM^B }\,,
}
through diagonalization of the right-handed fields described in Sec.\,\ref{sec:diag.1}.
Therefore $Y^d$ and $Y^B$ depend on common parameters which leads to correlations.
Note that $Y^d$ largely fixes the flavor structure of the SM $d$-sector if we ignore deviations from unitarity.
So, in generic VLQ models the mixing angles $\theta_L$ in \eqref{thetaL} are globally suppressed by the heavy VLQ masses but its flavor structure is completely free and dictated by $Y^B$, unrelated to the structure in $Y^d$.
That is not the case in Nelson-Barr type models where $Y^B$ is related to $Y^d$ by the sum rule \eqref{sum.rule} and then $\theta_L$ is correlated with $Y^d$.
We can explicitly write $\theta_L$ in terms of the parameters in the original Lagrangian \eqref{yuk:NB} as
\eq{
\theta_L=\frac{v}{\sqrt{2}}\cY^d {\cM^{Bd}}^\dag H_B^{-1}\,,
}
by using eqs. \eqref{Yd:YB}, \eqref{W:db;BB}, and \eqref{thetaL}.

\section{Number of CP violating phases}
\label{sec:CPV}

In Sec.\,\ref{sec:NB} we showed that a model of NB-VLQs contains one less parameter than a generic model of VLQs.
Among these parameters, it is interesting to know how many of them are CP violating.
The case of generic VLQs is well known\,\cite{branco:book}.
Here we will see that the case of NB-VLQs will have less CP violating parameters than a naive calculation shows.

We can review the case of generic VLQs.
For that, it is sufficient to analyze the case of the CP conserving case and subtract these CP even parameters from \eqref{N.para:VLQ}.
CP conservation requires real parameters in the Lagrangians \eqref{yuk:SM} and \eqref{yuk:VLQ}.
In the basis where $M^B$ and $Y^u$ are diagonal we can count
\eq{
\label{N.param:VLQ:real}
N^{\rm real}_{param}=\ums{2}N_f(N_f+3)+n_B(N_f+1)\,
}
CP even parameters.
So the number of CP violating phases is the difference between \eqref{N.para:VLQ} and 
\eqref{N.param:VLQ:real}:
\eq{
N^{\rm phases}_{\rm param}=N_{\rm param}-N^{\rm real}_{\rm param}
=\frac{(N_f-1)(N_f-2)}{2}+n_B(N_f-1)\,.
}
For $n_B=0$ we obtain the usual Kobayashi-Maskawa result\,\cite{km}.
For $n_B=1$, the known result of two additional phases is recovered\,\cite{lavoura.branco,branco:book}.

We can do an analogous analysis for the case of NB-VLQs which will lead to the wrong number.
The total number of parameters was given in \eqref{N.param:NB}.
If the CP symmetry is defined in the usual form for the fields in \eqref{yuk:NB}, 
the CP conserving limit is achieved by taking real $\cM^{Bd}$ and all parameters of \eqref{yuk:NB} real.
This is equivalent to considering real parameters in the usual VLQ case and the number of parameters is the same as in eq.\,\eqref{N.param:VLQ:real}.
Comparing to the number of parameters in \eqref{N.param:NB}, the number of (soft) CP violating phases is
\eq{
N^{\rm phases}_{\rm param}\big|_{\rm NB}=(N_{\rm param}-N^{\rm real}_{\rm param})\big|_{\rm NB}
=n_B(N_f-1)\,.
}
So it appears that, for one NB-VLQ, there are two CP violating phases, one accounting for the CKM phase and the other being a new source.

The previous counting does not lead to the correct number of phases because of two ingredients.
The first is the rephasing freedom
\eq{
\label{rephasing:B}
B_{rL}\to e^{i\theta_r}B_{rL}\,,\quad
B_{rR}\to e^{i\theta_r}B_{rR}\,,
}
valid in the basis where $\cM^B$ in \eqref{yuk:NB} is diagonal.
This leads to rephasing from the left of $\cM^{Bd}$.
So if the phases of $\cM^{Bd}$ can be removed this way, there can be no CP violation.
And of course the CP symmetry needs to be redefined.

The second ingredient is that some parameters can be transferred from $\cM^{Bd}$ to $\cY^d$ if we perform a real orthogonal transformation $d_{iR}\to (O_{d_R})_{ij}d_{jR}$ in \eqref{yuk:NB} which induces
\eq{
\label{reparm:OdR}
\cY^d\to \cY^d O_{d_R}\,,\quad
\cM^{Bd}\to \cM^{Bd}O_{d_R}\,.
}
We can concentrate on the first row of $\cM^{Bd}$, $w_i=\cM^{Bd}_{1i}$ and
we specialize already to $N_f=3$.
If we only use the rephasing \eqref{rephasing:B}, we can only remove one phase of $w$ and two  phases remain. 
Instead, we can choose $\theta_1$ in \eqref{rephasing:B} so that the vectors $\re(w)$ and $\im(w)$ are orthogonal.\footnote{%
This is always achievable: use rephasing \eqref{w:B-number} to make $w{\cdot} w=\re(w){\cdot}\re(w)-\im(w){\cdot}\im(w)+2i\re(w){\cdot}\im(w)$ real.
Hence $\re(w)$ and $\im(w)$ would be orthogonal real 3-vectors.
}
Then the matrix $O_{d_R}$ can be further chosen such that 
\eq{
\label{w:ba}
w_i=\cM^{Bd}_{1i}\sim (0,ib,a)\,,
}
where $a,b$ are real positive.
So there is only one CP odd quantity.

In this basis, $O_{d_R}$ is fixed, except for discrete choices, and the rephasing in $B_1$ can no longer be applied. The rephasing of $B_{i}$, $i\ge 2$, remain.
Each of these rephasing transformations can remove one phase from each row of $\cM^{Bd}$.
We are left with 
\eq{
\label{N.cp.odd:true}
N^{\rm phases}_{\rm param}\big|_{\rm NB}=1+(n_B-1)\times 2\,,
}
CP odd quantities. Note that $N_f=3$.
Hence, for a single NB-VLQ, there is \emph{only one} CP odd quantity responsible for all CP violating effects.\,\footnote{%
We should make a distinction between this number and the number of phases in the BBP model\,\cite{BBP} where only one scalar was responsible for spontaneous CP violation.
Here, even if more scalars are present, effectively, only one phase is transmitted to the SM for a single NB-VLQ.
}
We will see this specific case in more detail in Sec.\,\ref{sec:param}.

One last comment is in order. It seems that the number \eqref{N.cp.odd:true} of CP odd quantities is in contradiction either with the number \eqref{N.param:VLQ:real} of CP even parameters in the CP conserving case or with the total number of parameters in \eqref{N.param:NB}.\footnote{%
We thank the anonymous referee for this observation.}
For example, for $n_B=1$, Eqs.\,\eqref{N.cp.odd:true}, \eqref{N.param:NB} and \eqref{N.param:VLQ:real} 
gives $1, 15$ and $13$ for the number of CP odd quantities in the NB case, the total number of parameters in the NB case, and the total number of parameters in the CP conserving case, respectively.
We discuss this apparent contradiction in appendix \ref{ap:contradiction} and illustrate using the $n_B=1$ case that the apparent contradiction is solved by realizing that in the CP conserving limit one CP \emph{even} parameter becomes unphysical.

\section{Irreducible flavor violation}
\label{sec:irred.flavor}

It is well known that the presence of VLQs induce new effects such as unitarity violation of the CKM matrix or flavor changing interactions mediated by the $Z$.
The first effect appears in the deviation from unitarity of the $3\times 3$ block of \eqref{V} involving the mixing of the known quarks
where the deviation is quantified by $\delta X^d$ in \eqref{Theta}.
The same quantity induces flavor nonuniversal (diagonal) and flavor violating (off-diagonal) interactions of usual $d$ quarks mediated by the $Z$ in the upper-left $3\times 3$ block of $X^d$ in \eqref{X^d}.
The flavor violating part is traditionally called flavor changing neutral currents (FCNC).

We will show that for VLQs of Nelson-Barr type, unlike generic VLQs, the quantity $\delta X^d$ cannot be switched off, it cannot be diagonal, and thus an \textit{irreducible} amount of flavor violation mediated by the $Z$ boson is always present.
That $\theta_L$ cannot vanish is understood because if the heavy VLQs decouple, they would not transmit the required CP violation to the SM.
The situation here is stronger: the new interactions are necessarily flavor violating.

To understand such a flavor violation, we use the sum rule \eqref{sum.rule} which can be rewritten as 
\eq{
\label{flavor.vio}
V_{d_L}^\dag\frac{2}{v^2}\theta_LH_B\theta_L^\dag V_{d_L}
=V_{d_L}^\dag\cY^d{\cY^d}^\tp V_{d_L}-(\hY^d)^2\,.
}
Recall that $V_{d_L}$ diagonalizes $Y^d{Y^d}^\dag$ and $\hY^d$ denotes the diagonalized version of $Y^d$ which should approximately match the square root of the known values in \eqref{ckm:sm}.
Now, since the left-hand side of \eqref{flavor.vio} is positive semidefinite\,\footnote{%
A matrix $\mathcal{A}$ is positive definite (semidefinite) if $x^{\dag}\mathcal{A}x>0$ ($x^{\dag}\mathcal{A}x\ge 0$) for all vectors $x\neq 0$.} 
(positive definite for $n_B\ge 3$), so should be the right-hand side.
However, since $\cY^d{\cY^d}^\tp$ is real symmetric, it cannot be diagonalized by the intrinsically complex matrix $V_{d_L}\approx \ckmsm$. So the cancellation in the right-hand side of \eqref{flavor.vio} cannot be complete and it \emph{must have nonzero off-diagonal entries}.
This immediately translates into non-zero off-diagonal entries in the left-hand side of \eqref{flavor.vio} which is related to the flavor violation in $\delta X^d$.
Generically, we expect that if \eqref{flavor.vio} is nondiagonal, then $\delta X^d$ is also nondiagonal and flavor violating.
For a single VLQ, the implication is directly ensured.

Now we should emphasize an important difference between the mixing matrix of the present model,
$V_{d_L}$ defined in \eqref{ckm:sm}, and the CKM matrix $\ckmsm$ of the SM:
phases of  $V_{d_L}$ at the left are physical and cannot be removed by rephasing transformations of up-type $u_{iL}$ fields.
The reason is that removing phases of the mixing $V_{ij}$ of usual quarks in the Lagrangian \eqref{L:W} reintroduces the same phases in the mixing $V_{i,3+a}$ with $B_{aL}$.
So we should parametrize
\eq{
\label{majora-like}
V_{d_L}=\mtrx{1&&\cr &e^{i\beta_2}&\cr &&e^{i\beta_3}}\ckmsm\,,
}
where $\ckmsm$ is the CKM matrix of the SM with some fixed rephasing convention.
The phases $\beta_2,\beta_3$ are new free parameters.

To quantify the minimal irreducible flavor violation that might be present, we can minimize the right-hand side of \eqref{flavor.vio} using some kind of norm, restricted by the constraint that it should be positive semidefinite.\,\footnote{%
The notion of minimal irreducible flavor violation is not uniquely defined because it depends on the quantity to be minimized.
}
This exercise is performed in appendix \ref{ap:min} and we find that the right-hand side of \eqref{flavor.vio} is at most of the order of $10^{-7}$.
Therefore, although undetectably small, the presence of NB-VLQs introduces an irreducible amount of FCNC that cannot be reduced to zero even by fine-tuning.
In the next section, we will make a more quantitative study for the simplest but still intricate case of a single NB-VLQ.

\section{Seesaw parametrization for a single NB-VLQ}
\label{sec:param}

We focus here on the case of a single NB-VLQ ($n_B=1$).
To quantitatively test the model, we would like to parametrize the 15 physical parameters of the Lagrangian \eqref{yuk:NB} keeping fixed 10 relations that should account for the 10 parameters of the SM flavor sector. Five free parameters remain to describe BSM physics.
Within the seesaw approximation, an explicit and analytical parametrization will be shown below. We call this parametrization the \emph{seesaw parametrization}.

Three parameters just correspond to the up-type quark yukawas (or masses), 
$\hat{\cY}^u=\sqrt{2}v^{-1}\diag(m_u,m_c,m_t)$.
So we need to parametrize the quantities
\eq{
\label{NB:params}
\{\cY^d,\cM^{Bd},\cM^B\}
}
using 12 parameters among which 7 should be fixed to account for the SM down sector Yukawas and CKM mixing.

Among the fundamental parameters in the Lagrangian \eqref{yuk:NB}, $\cM^B$ is just a real number and can be traded by the mass of the heavy quark
in \eqref{HB},
\eq{
\label{mB}
H_B=\cM^{Bd}{\cM^{Bd}}^\dag+(\cM^B)^2 =M_B^2\,.
}
We use $M_{B}=M_{B_1}$ to denote \emph{one} VLQ mass instead of $M^B$ which is reserved for the multiparticle mass matrix.
The matrix $\cM^{Bd}\sim 1\times 3$ is complex and can be parametrized by a complex vector $w$ by
\eq{
\label{w}
{\cM^{Bd}}^\dag=M_B\,w=M_B(w_1,w_2,w_3)^\tp\,.
}
The relation \eqref{mB} means that 
\eq{
\label{w<1}
0<|w|< 1\,.
}
The border values $|w|=0$ or $|w|=1$ are respectively excluded because, according to \eqref{Yd:2}, the CKM matrix would be real ($B$ decouples) or one of the SM quarks would be massless.
One parameter in $w$ (in $\cM^{Bd}$) is unphysical because it can be removed by $B$-number conservation in the basis \eqref{yuk:NB}, i.e.,
\eq{
\label{w:B-number}
w\to e^{i\alpha}w\,.
}
is innocuous.
We are left with 11 parameters in $\{\cY^d,w\}$.

We note that some parameters can be transferred from $w$ to $\cY^d$ if we perform a real orthogonal transformation inducing \eqref{reparm:OdR}.
See discussion in Sec.\,\ref{sec:CPV}.
So we can choose $O_{d_R}$ and $\alpha$ such that
\eq{
\label{w:a,b}
w=\mtrx{0\cr ib\cr a}\,,
}
with $a,b$ being real positive parameters subjected to $a^2+b^2<1$; cf.\,\eqref{w<1}.
Moreover, we can choose $b\le a$ because the roles of $a$ and $b$ can be reversed due to the rephasing freedom \eqref{w:B-number} and reparametrization freedom \eqref{reparm:OdR}.
Incidentally, both $a,b$ need to be nonzero for CP violation.
Since a special form for $w$ was chosen, $\cY^d$ needs to be a generic real $3\times 3$ matrix with 9 parameters.
The structure \eqref{w:a,b} means $\cM^{Bd}_1=0$ and $d_{1R}$ in \eqref{yuk:NB} couples only through $\cY^d_{i1}$ so that, e.g., $\cY^d_{11}>0$ can be conventionally chosen.
The total number matches 11 parameters.
At this point, it is easy to see that CP will be conserved in the limit $b\to 0$, in accordance with the discussion of Sec.\,\ref{sec:CPV}, i.e., $b$ is the only CP odd quantity of the model.

Now we need to write $\cY^d$ in terms of the 7 parameters in the down-sector Yukawa matrix of SM in the basis where the up-sector Yukawa is diagonal: three down-sector yukawa couplings and the four parameters in the CKM matrix.
This inversion process will involve trading several parameters in favor of others.

Within the seesaw approximation \eqref{seesaw}, the down-sector Yukawa matrix is given by \eqref{Yd:2}.
So we need to solve for $\cY^d$ and $w$ in
\eq{
\label{Yd:nB=1}
\cY^d\big(\id_3-ww^\dag\big){\cY^d}^\tp
=Y^d{Y^d}^\dag
=V_{d_L}\big(\hY^d\big)^2 V_{d_L}^\dag
\,,
}
where $\hY^d=\sqrt{2}v^{-1}\diag(m_d,m_s,m_b)$ is the SM Yukawa couplings and $V_{d_L}$ is the CKM matrix of the SM in the standard parametrization with the addition of the two phases $\beta_2,\beta_3$ in \eqref{majora-like}.
Considering the nontrivial phase in the CKM matrix of the SM, irrespective of $\beta_2,\beta_3$, the right-hand side of the last equality in \eqref{Yd:nB=1} is essentially complex, and then the CP conserving limit is no longer possible once we choose to describe the SM with CP violation.
The soft CP violation should account for the CP violation of the SM.

We can rewrite the previous equation as 
\eq{
{\cY^d}^{-1}Y^d{Y^d}^\dag{\cY^d}^{\tp-1}=\big(\id_3-ww^\dag\big)\,.
}
In the basis \eqref{w:a,b}, the real and imaginary parts must obey
\subeqali[toinvert]{
\label{toinvert:re}
{\cY^d}^{-1}\re(Y^d{Y^d}^\dag){\cY^d}^{\tp-1}&=\mtrx{1&&\cr &1-b^2& \cr & &1-a^2}\,,
\\
\label{toinvert:im}
{\cY^d}^{-1}\im(Y^d{Y^d}^\dag){\cY^d}^{\tp-1}&=ab\mtrx{0&&\cr &0& -1\cr & 1& 0}\,.
}

Given that the real part of a hermitian positive definite matrix is also positive definite, 
we can define the real symmetric matrix
\eq{
\label{def:A}
A\equiv \Big(\re(Y^d{Y^d}^\dag)\Big)^{1/2}\,,
}
which is uniquely defined for each value of $\beta_2,\beta_3$.
The left-hand side of \eqref{toinvert:re} is consequently positive definite.
The right-hand side is also positive definite due to \eqref{w<1}.
So we define
\eq{
B\equiv \diag(1,\sqrt{1-b^2},\sqrt{1-a^2})\,.
}
Then \eqref{toinvert:re} is only possible if 
\eq{
{\cY^d}^{-1}A \cO=B\,,
}
where $\cO$ is a real orthogonal matrix.
Inverting the relation, we have the solution
\eq{
\label{cal.Yd:inverse}
\cY^d=A\cO B^{-1}\,.
}

Plugging this solution to \eqref{toinvert:im}, we obtain
\eq{
\label{cano:Im}
\cO^\tp A^{-1}\im(Y^d{Y^d}^\dag)A^{-1}\cO=\frac{a}{\sqrt{1-a^2}}\frac{b}{\sqrt{1-b^2}}
\mtrx{0&&\cr &0& -1\cr & 1& 0}
\,.
}
Then $\cO$ is the matrix that transforms the real antisymmetric matrix $C\equiv A^{-1}\im(Y^d{Y^d}^\dag)A^{-1}$ to the canonical form
\eq{
\label{def:C}
C\sim \mu\mtrx{0 &&\cr &0&-1\cr &1&0}\,.
}
$\mu>0$ is uniquely determined, for example, because $C$
should have eigenvalues $(0,i\mu,-i\mu)$.
In appendix \ref{ap:mu} we show that $\mu<1$ as well.
Then $a$ and $b$ are not independent but related by
\eq{
\label{def:mu}
\frac{a}{\sqrt{1-a^2}}\frac{b}{\sqrt{1-b^2}}=\mu\,.
}

The matrix $\cO$ is formed by unit column vectors,
\eq{
\cO=\left(\begin{array}{c|c|c}
e_1& e_2 & e_3
\end{array}
\right)\,,
}
such that $e_1$ is the only real eigenvector of $C$ with zero eigenvalue while $e_2,e_3$ generate the space orthogonal to $e_1$, with the relative sign between $e_2,e_3$ determined by \eqref{cano:Im}.

We can fix $e_2,e_3$ using some convention. For example, we can take the real and imaginary part of the complex eigenvector of $C$ associated to $i\mu$.
Then there is one degree of freedom associated to the rotation in the plane $e_2{-}e_3$ which we can parametrize as
\eq{
\cO=\left(\begin{array}{c|c|c}
e_1& e_2 & e_3
\end{array}
\right)\mtrx{1&&\cr &\cos\gamma &\sin\gamma\cr &-\sin\gamma&\cos\gamma}\,.
}

Now we can check the number of parameters.
The four parameters that are free are 
\eq{
\label{seesaw:param}
\{\beta_2,\beta_3,\gamma,b\}\,.
}
Other 7 parameters accounts for the three down quark Yukawa couplings, three CKM mixing angles and one Dirac CP phase.
The total is 11. If we include $M_B$, we get the necessary 12 parameters.

So we succeeded in parametrizing the theory with new free parameters keeping the SM flavor parameters compatible within the seesaw approximation.
This is the \emph{seesaw parametrization}.

We note that the phases $\beta_2,\beta_3$ appear as parameters only as a result of the inversion process in order to keep the SM Yukawa fixed in \eqref{Yd:nB=1}.
In the initial set of parameters \eqref{NB:params}, there is only one CP-odd quantity $b$ as discussed previously.
The inversion process also introduces an implicit $b$ dependence in $\cY^d$ through \eqref{cal.Yd:inverse}.
Analogously, the $a$ parameter is not a free parameter anymore, being fixed by \eqref{def:mu}.

At this point, all quantities can be expressed in terms of the parameters of the SM and the parameters in \eqref{seesaw:param}.
For example, the matrix $\Theta$ in \eqref{V} which describes the coupling of the VLQ with $W$ and $u_{iL}$ is given by 
\eq{
\label{Theta:NB}
\Theta=\frac{v}{\sqrt{2}M_B}V_{d_L}^\dag A\cO 
\mtrx{0\cr \displaystyle\frac{ib}{\sqrt{1-b^2}}\cr \displaystyle\frac{a}{\sqrt{1-a^2}}}
\,.
}
This matrix also dictates the FCNC to the $Z$.
The matrix $A$ defined in \eqref{def:A} is clearly hierarchical and this hierarchy is inherited by $\Theta$.
We show in Sec.\,\ref{sec:pheno} a plot of the quantity $|V_{iB}|$, which is rotated by CKM mixing and equally hierarchical.

We should emphasize that the hierarchical structure of \eqref{Theta:NB} is not a generic feature of a theory with a VLQ. The feature arises for one NB-VLQ once the SM Yukawa is reproduced and the matrix $A$ depends on the $d$-quark mass matrix of the SM which is hierarchical.
Instead, for a generic VLQ, $\Theta$ depends on $Y^B$ which does not need to have a structure similar to $Y^d$, although some parameters are strongly constrained from phenomenology.

\begin{figure}[h]
\includegraphics[scale=0.38]{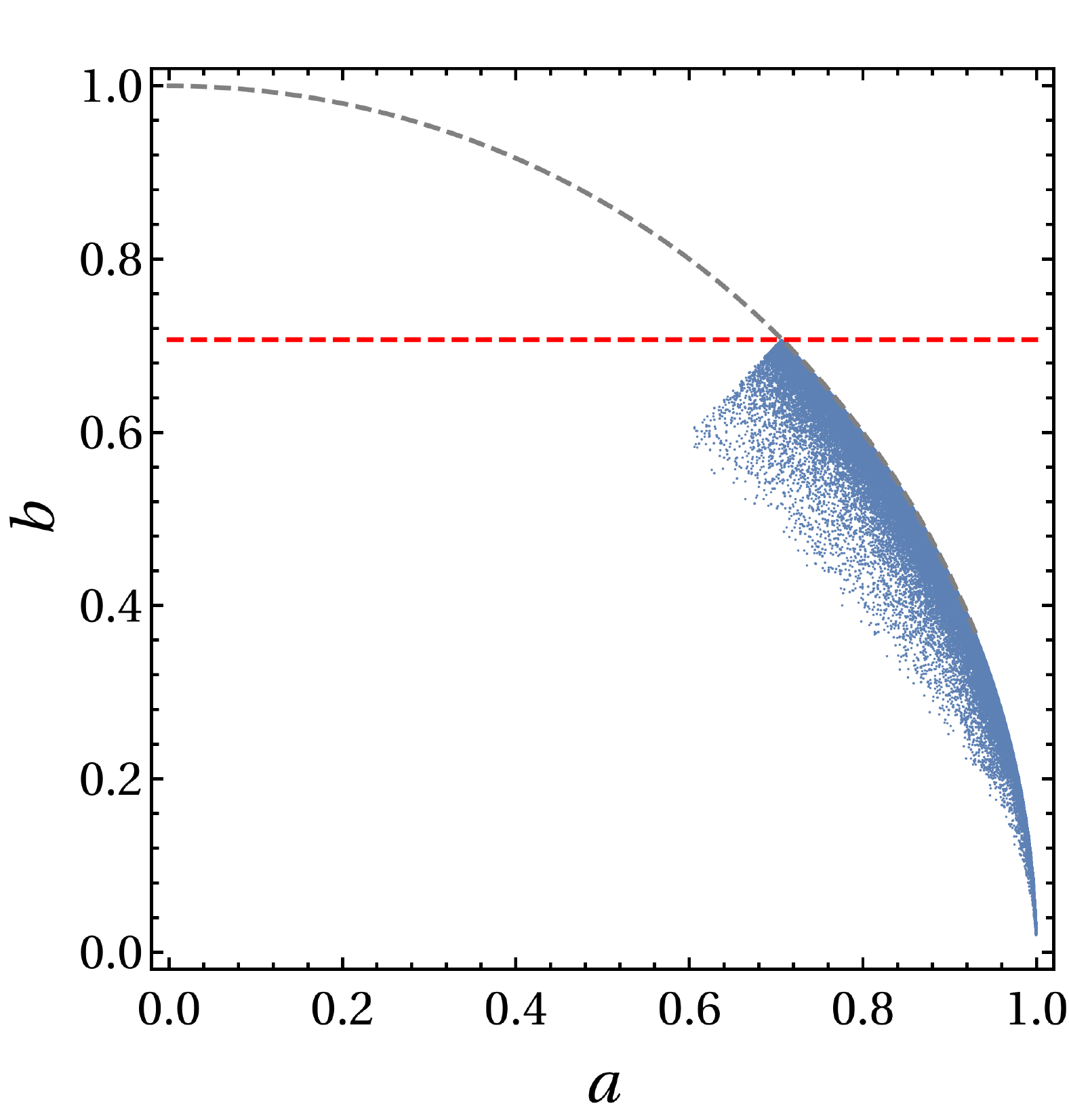}
\caption{
Distribution of $(a,b)$ for $b\in [0,1/\sqrt{2}]$ with $b\le a$. The gray dashed line represents $a^2+b^2=1$. The red dashed line represents $b=1/\sqrt{2}$.
}
\label{fig:ab}
\end{figure}

Obviously the seesaw approximation is not valid everywhere: the masses coming from the explicit diagonalization of \eqref{diag.2} compared to the ones coming from the seesaw approximation in 
\eqref{seesaw.block.1} and \eqref{HB} might deviate.
This deviation is potentially larger when $b$ is very small.
We have checked that $(a,b)$ is confined approximately to the unit circle so that $a^2+b^2\approx 1$; see Fig.\,\ref{fig:ab}.
Remember that $b\le a$.
And the approximation to the unit circle is better when $b$ is very small.
So $a\approx 1$ when $b\ll 1$.
Considering that $a^2+b^2$ is at most unity, we can consider $b\le 1/\sqrt{2}$.
This property explains the possible enhancement of the mixing of the VLQ with the SM quarks when $b\approx 0$. In this case, \eqref{Theta:NB} contains a term involving $1/\sqrt{1-a^2}$ which is enhanced.

\begin{figure}[h]
\includegraphics[scale=0.45]{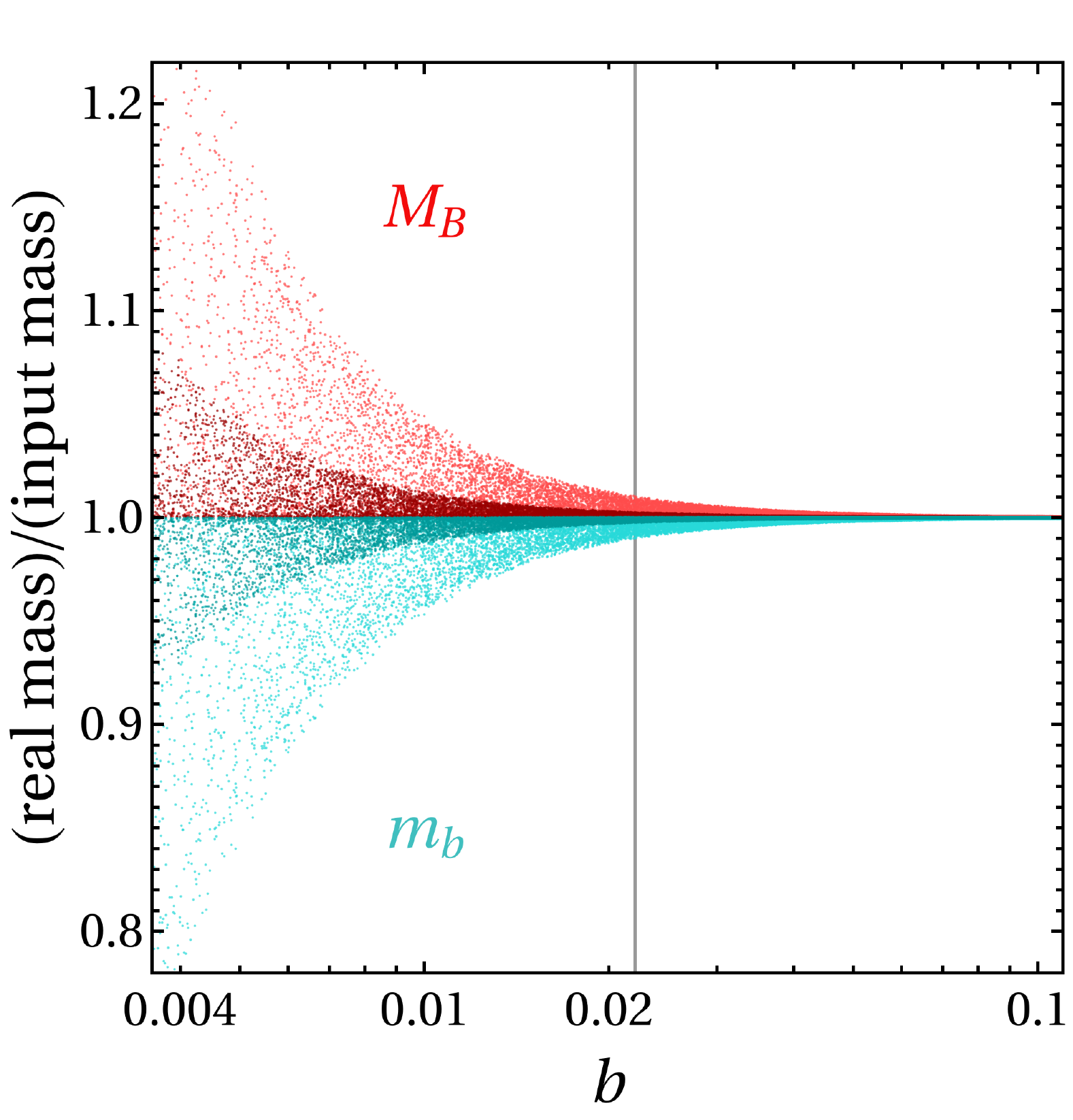}
\caption{
The ratio between the real mass and the input mass for small $b$.
The gray vertical line marks the left cut in \eqref{b:interval}.
The lighter colors use $M_B=1.3\,\unit{TeV}$ whereas the darker colors use $M_B=2.6\,\unit{TeV}$.
}
\label{fig:mass.ratio}
\end{figure}

The deviation for $m_b$ (downward) and $M_B$ (upward) can be seen in Fig.\,\ref{fig:mass.ratio} for small $b$ where we show the ratio between the real mass and the seesaw mass or input mass.
The deviation for the other quark masses are much smaller.
For definiteness we use $M_B=1.3\,\unit{TeV}$ and the seesaw structure tell us that larger $M_B$ will lead to smaller deviations.
We can see that for values of $b$ smaller than $0.0221$ (gray line), 
the deviation for $m_b$ gets larger than 1\%, which is roughly the error for $m_b$ in the SM\,\cite{PDG}.
For those values of $b$, the deviation of the CKM matrix \eqref{V} calculated exactly (second equality) compared to the one calculated using the leading seesaw approximation (third equality) can be also seen to be roughly below 1\%. The larger deviation being on the $|V_{td}|$ element which reaches $1.2\%$ when $b=0.0221$.
Therefore, for $M_B=1.3\,\unit{TeV}$, the seesaw parametrization will be reliable roughly within 1\% in the interval
\eq{
\label{b:interval}
b|_{ss}\in \big[0.022,1/\sqrt{2}\big]\,.
}
For comparison, Fig.\,\ref{fig:mass.ratio} also shows in darker points the deviation for $M_B=2.6\,\unit{TeV}$. One can see that the deviation is much smaller.

In Fig.\,\ref{fig:ab}, Fig.\,\ref{fig:mass.ratio} and subsequent Fig.\,\ref{fig:ViB}, we use 
$b\in [0,\sqrt{2}]$, and all $\{\gamma,\beta_{2},\beta_{3}\}$ in the whole allowed range of $[0,2\pi)$, while we take the d-quark masses according PDG\,\cite{PDG} and the best-fit values of angles and phase of the CKM matrix according to CKMfitter\,\cite{ckmfitter}. The values are listed in Sec.\,\ref{sec:pheno}.

Outside the range above, we can still use the inverting relation \eqref{cal.Yd:inverse} for $\cY^d$ and use $Y^d{Y^d}^\dag$ in \eqref{Yd:nB=1} as input.
Given that the input masses and the CKM matrix elements might deviate, we can try to compensate for such a deviation by changing the input values in $Y^d{Y^d}^\dag$ and $M_B$.
We will not treat this case any further and will concentrate on the seesaw parametrization.

\section{Phenomenology for $n_B=1$}
\label{sec:pheno}

\subsection{Hierarchical mixing}

\begin{figure}[h]
\includegraphics[scale=0.55]{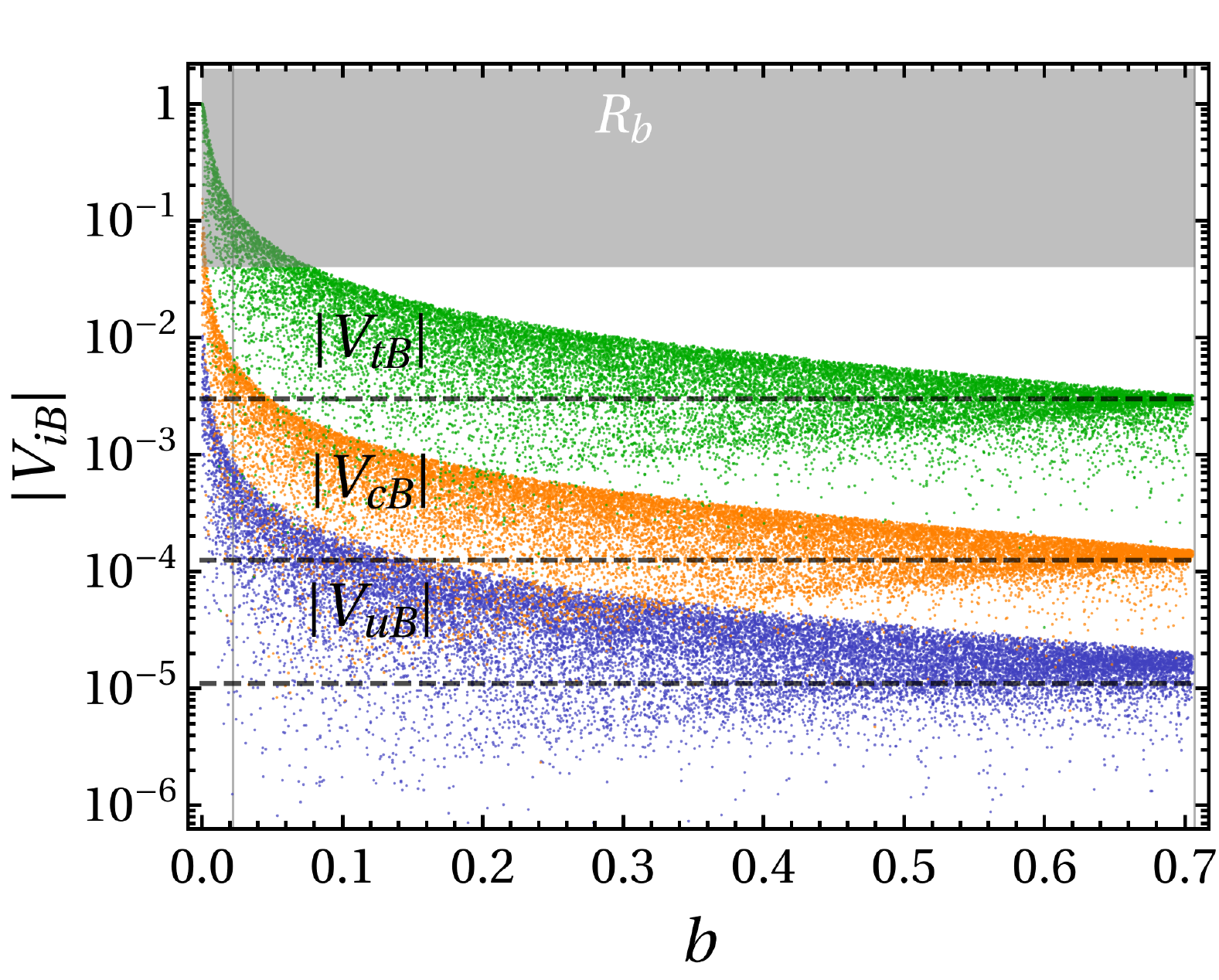}
\caption{\label{fig:ViB}
The CKM matrix elements $|V_{iB}|$ as a function of $b$.
The gray vertical lines mark the interval in \eqref{b:interval}.
The shaded area shows the values excluded from $R_b$\,\cite{saavedra:handbook}.
The black dashed lines shows the case of the VLQ mixing only with third family with $s_{\theta_L}=0.003$ in \eqref{ViB:3.fam}. 
See text for details.
}
\end{figure}

Using the seesaw  parametrization described in Sec.\,\ref{sec:param}, we can check that the couplings of the quark $B$ with the up-type quarks and the boson $W$ are hierarchical: $|V_{tB}|\gg |V_{cB}|\gg |V_{uB}|$.
This information is depicted in Fig.\,\ref{fig:ViB}. 
We use $M_B=1.3\,\unit{TeV}$, the SM down-type Yukawa couplings in $\hat{Y}^d$ and the CKM matrix of the SM in \eqref{majora-like} to define
$V_{d_L}$; the phases $\beta_2,\beta_3$ and the angle $\gamma$ are varied in the whole range of $[0,2\pi]$. 
The extraction of $V$, however, is done diagonalizing the $4\times 4$ mass matrix explicitly.
Note that the points for $b\ll 1$ may not correspond to physical points because the seesaw approximation is not reliable in such a regime.

Because $V_{iB}$ are hierarchical, the $B$ quark couples dominantly with the top and we can use the current constraint coming from direct searches at the LHC\,\cite{atlas,cms}:
\eq{
M_B\gtrsim 1.3\,\unit{TeV}\,.
}
So, when we use a fixed mass, we will use the lower limit $M_B=1.3\,\unit{TeV}$.

The information of $|V_{iB}|$ in Fig.\,\ref{fig:ViB} also roughly translates into $|X_{iB}|$ because
\eq{
|X_{iB}|=\left|\sum_{j=1}^{3}V^*_{ji}V_{jB}\right|\approx \left|\sum_{j=1}^{3}({\ckmsm}^*)_{ji}V_{jB}\right|
\approx |V_{iB}|\,,
}
ignoring the CKM mixing in the last approximation.

\subsection{Comparison with third family only mixing}
\label{sec:3rd.fam}

The frequently considered benchmark case where one VLQ couples solely with the third family of the SM also exhibits a hierarchical mixing of the VLQ with the SM quarks through $W$.
Here  we briefly analyze the difference between this case and the NB case with respect to the $3\times 4$ CKM matrix.

We can define the case of mixing with third family only by assuming in \eqref{diag.2} the structure
\eq{
Y^d=V_{d_L}\hat{Y}^d\,,\quad
Y^B=V_{d_L}\mtrx{0\cr0\cr y_B}\,,
}
as a particular case of the generic VLQ in the basis there $Y^u=\hat{Y}^u$.
In this special case, we have
\eq{
\label{U:3.fam}
U_{d_L}=
\left(\begin{array}{c|c}V_{d_L}& \cr\hline & 1\end{array}\right)
\left(\begin{array}{ccc|c}
1&&&\cr&1&&\cr &&c_{\theta_L}& s_{\theta_L}\cr\hline &&-s_{\theta_L}&c_{\theta_L}
\end{array}\right)\,,
}
and the $3\times 4$ CKM matrix $V$ is obtained by chopping the last row.
The shorthand $s_{\theta_L}$ denotes $\sin\theta_L$ as usual and the same is valid for the cosine.
The mixing angle can be calculated exactly and yields\,\cite{saavedra:handbook}
\eq{
\tan2\theta_L=\frac{\sqrt{2}|y_B|v M_B}{M_B^2-|\hat{Y}^d_{33}|^2v^2/2-|y_B|^2v^2/2}\,,
}
where the analogous angle on the right-handed quarks is further suppressed:
\eq{
\tan\theta_R=\frac{m_b}{M_B}\tan\theta_L\,.
}
These angles match \eqref{thetaL} and \eqref{thetaR} within the seesaw approximation.

In this case the entries $V_{uB}$ and $V_{cB}$ are not strictly zero, but they are suppressed by the SM CKM compared to $V_{tB}$ following the relation
\eq{
\label{ViB:3.fam}
|V_{iB}|=s_{\theta_L}|V_{ib}|,\quad i=u,c,t,
}
where $s_{\theta_L}\approx v|y_B|/\sqrt{2}M_B$.
These $|V_{tB}|,|V_{cB}|,|V_{uB}|$ are shown in dashed lines in Fig.\,\ref{fig:ViB} with $s_{\theta_L}=0.003$, corresponding to $v|y_B|/\sqrt{2}\approx 3.9\,\unit{GeV}$ for $M_B=1.3\,\unit{TeV}$.
We can see that $|V_{iB}|$ in the NB-VLQ case roughly follow \eqref{ViB:3.fam} for $b=1/\sqrt{2}$ but $|V_{uB}|$ tends to be larger. For smaller $b$, the deviations from these proportions are much larger.

In contrast to the coupling with the $W$, the FCNC with the $Z$ is only present between $bB$ as
\eq{
\label{Xd:3rd.fam}
X^d=V^\dag V=\mtrx{1&&&\cr &1&&\cr &&c^2_{\theta_L}&s_{\theta_L}c_{\theta_L}\cr &&s_{\theta_L}c_{\theta_L}&s^2_{\theta_L}}\,.
}

\subsection{Flavor constraints}
\label{sec:flavor}

Here we will show that, due to the hierarchical structure of the CKM matrix, the case of one NB-VLQ is largely \textit{flavor safe} in the regime where the seesaw parametrization holds. 
With the seesaw parametrization devised in Sec.\,\ref{sec:param}, 
the parameters of the SM such as quark masses and $|V_{ij}|$, can be \textit{chosen} as input to be as close to the experimental value as desired, within 1\%.
So constraints coming from them are easily avoided at the parametrization stage.
Here we will focus on the possible constraints on the model and will not try to map the detailed available parameter space. The latter would require a dedicated global fit procedure because small deviations of some parameters of the SM are possible in the presence of the VLQ\,\cite{saavedra:flavor, buras.celis}.

Most of the moduli $|V_{ij}|$ of the CKM matrix in the SM are extracted from tree level processes.
The exceptions are $|V_{td}|$ and $|V_{ts}|$ which are extracted from $B_d^0$ and $B_s^0$ meson oscillations through box diagrams involving the top.
These processes may receive contributions from the VLQ and therefore
$|V_{td}|$ and $|V_{ts}|$ may deviate slightly from the SM values.

The experimental values for $|V_{ij}|$ at 1$\sigma$ are\,\cite{PDG}:
\eq{
\label{|V|:exp}
|V_{ij}|_{\rm exp}=
\mtrx{
0.97420\pm 0.00021 & 0.2243\pm 0.0005 & (3.94\pm 0.36)\times 10^{-3}
\cr
0.218\pm 0.004 & 0.997\pm 0.017 & (42.2\pm 0.8)\times 10^{-3}
\cr
\underline{(8.1\pm 0.5)\times 10^{-3}} & \underline{(39.4\pm 2.3)\times 10^{-3}} & 1.019\pm 0.025
}\,.
}
For consistency, one needs to consider these constraints at 2$\sigma$ because 
the values above are not consistent with a unitary CKM matrix.
The values for $|V_{td}|$ and $|V_{ts}|$ are underlined to emphasize that they are not extracted from tree level processes.

For comparison, we can also show the values for the magnitudes of the CKM elements obtained from the combination of the various experiments and assuming unitarity.
The result of CKMfitter\,\cite{ckmfitter} is
\eq{
\label{|V|:exp:fit} 
|V_{ij}|_{\rm exp}^{\rm fit}=
\mtrx{
0.974390^{+0.000014}_{-0.000058} & 0.224834^{+0.000252}_{-0.000059} & 0.003683^{+0.000075}_{-0.000061}
\cr
0.224701^{+0.000254}_{-0.000058} & 0.973539^{+0.000038}_{-0.000060} & 0.04162^{+0.00026}_{-0.00080}
\cr
\underline{0.008545^{+0.000075}_{-0.000157}} & \underline{0.04090^{+0.00026}_{-0.00076}} & 0.999127^{+0.000032}_{-0.000012}
}
\,.
}
Taking the best-fit point, we can extract 
\eq{
\label{ckm:bf}
\theta_{12}=0.226776,\quad
\theta_{23}=0.04164,\quad
\theta_{13}=0.003680,\quad
\delta=1.149\,,
}
for the standard parametrization; see eq.\,\eqref{ckm:standard} below.
These values will be used as input in most places.
We also list the central values for the down quark masses\,\cite{PDG}:
\eq{
\label{d-quark.masses}
\bar{m}_d=4.67\,\unit{MeV}\,,\quad
\bar{m}_s=93\,\unit{MeV}\,,\quad
\bar{m}_b=4.18\,\unit{GeV}\,.
}
These are $\overline{\text{MS}}$ masses for which the first two are determined at $\mu=2\,\unit{GeV}$ while $\bar{m}_b$ is at $\mu=\bar{m}_b$. We ignore the running between these two scales.

Since the $3\times 3$ block $|V_{ij}|$ of the CKM matrix can be as close to the SM values as desired, within $1\%$, and there is no clear tension of flavor data with the values \eqref{|V|:exp:fit} of SM\,\cite{PDG}, there is basically no constraint involving them.
Also, since $|V_{iB}|$ are hierarchical, many of the constraints on them are indistinguishable from the case discussed in Sec.\,\ref{sec:3rd.fam} of mixing with the third family only.
In the latter case, the strongest constraint comes from\,\cite{saavedra:handbook}
\eq{
R_b=\frac{\Gamma(Z\to b\bar{b})}{\Gamma(Z\to\text{hadrons})}\,,
}
which depends strongly on $X^d_{bb}$.
Such a constraint translates into $\sin\theta_L\le 0.04$ for the 
case of mixing only with the third family; cf.\,\eqref{Xd:3rd.fam}.
Considering the CKM mixing is hierarchical, we simply impose
\eq{
\label{Rb:VtB}
|V_{tB}|< 0.04\,.
}
We show this constraint as a shaded area in Fig.\,\ref{fig:ViB}.
We can see that only a small portion of the points, corresponding to small $b$, are excluded.

\begin{figure}[h!]
\hspace*{-2em}
\includegraphics[scale=0.55]{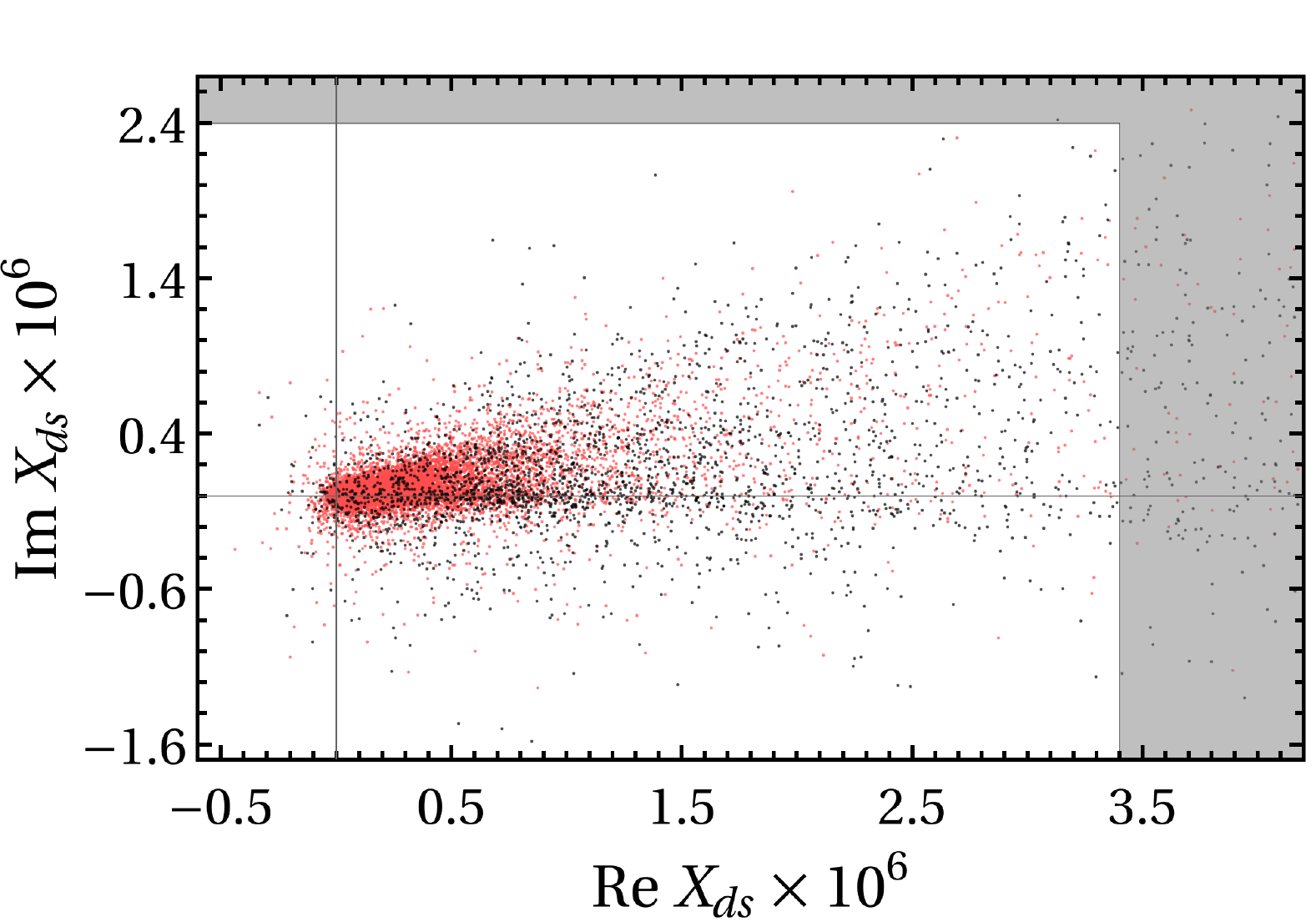}
\caption{\label{fig:ReX} 
Scatter plot (red and black) for the real and imaginary parts of $X_{ds}$ for the NB-VLQ model. In black we show the points excluded from $R_b$; cf.\ \eqref{Rb:VtB}.
The shaded area is excluded by \eqref{constraint:reX}.
}
\end{figure}

Less importantly than the constraint from $R_b$, CP violation in the kaon system is able to further exclude some points.
Ref.\,\cite{saavedra:flavor} reports the following constraints on $\re(X_{ds})$ (mainly from $K_L\to \mu^+\mu^-$) and $\im(X_{ds})$ (mainly from $\epsilon'/\epsilon$):
\eq{
\label{constraint:reX}
\re(X_{ds})\in \big[-1.0\times 10^{-5},3.4\times 10^{-6}\big]\,,
\quad
\im(X_{ds})\in \big[-2.7\times 10^{-6},2.4\times 10^{-6}\big]\,.
}
We have checked that the related constraint on $\im(Y^B_dY^{B*}_s)$ shown in Ref.\,\cite{buras.celis} is easily passed as well.
To impose these constraints on $X_{ds}$, we use the convention where $V_{ud}$ and $V_{us}$ are real.
These constraints are shown in Fig.\,\ref{fig:ReX} as shaded areas on top of the scatter plot for $\re(X_{ds})$ against $\im(X_{ds})$ in the NB-VLQ model.
These points are generated using the seesaw parametrization 
with $b$ in the interval \eqref{b:interval} and with the Yukawa $Y^d$ of the SM as input. This Yukawa can be recovered from the quark masses \eqref{d-quark.masses} and the SM CKM with the best-fit values \eqref{ckm:bf}. The rest of the parameters, $\beta_2,\beta_3,\gamma$, are varied in their whole possible range.
The red points pass the constraint from $R_b$ while the black ones do not. We can see that the constraint from $\re(X_{ds})$ is more important than from $\im(X_{ds})$ in the model.

\begin{figure}[h!]
\hspace*{-2em}
\includegraphics[scale=1.1]{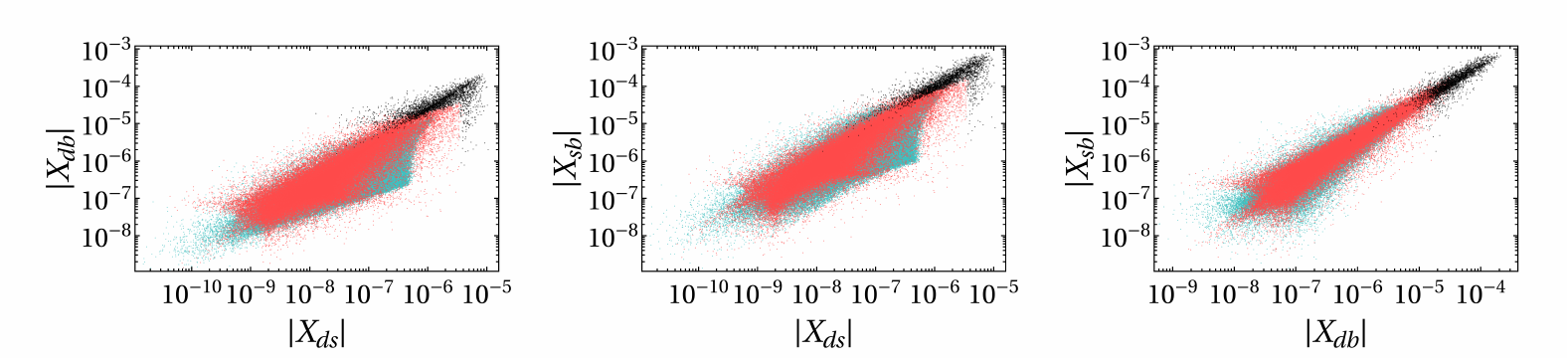}
\caption{\label{fig:X} Correlations among different elements of $|X^d_{ij}|$, for NB-VLQs (red) or generic VLQs (blue) using the angles \eqref{thetai4:range}. 
In black we show the points excluded from $R_b$ and CP violation in the kaon sistem; cf.\ \eqref{Rb:VtB} and \eqref{constraint:reX}.}
\end{figure}

The remaining flavor constraints are easily satisfied.
One can see that in Fig.\,\ref{fig:X} where the possible values for $|X^{d}_{ij}|$ for $(ij)=(ds),(db),(sb)$ are shown in red points.
We see that hierarchical $|V_{iB}|$ translates into a strong correlation among them.
The points excluded by \eqref{Rb:VtB} and \eqref{constraint:reX} are marked in black and we can see that they correspond to the largest values for $|X^d_{ij}|$.
The remaining points are easily compatible with other flavor constraints \cite{saavedra:flavor,buras.celis}.

\subsection{Comparison with generic VLQs}

Here we compare the hierarchical structure of the CKM matrix for the VLQ of Nelson-Barr type with the generic case.
We will see that the hierarchy in $X^d$ can be emulated by choosing an appropriate parametrization and hierarchical angles for the angles beyond the SM.

We use the parametrization\,\cite{gouvea}
\eq{
\label{V:param:angles}
U=
	\left(
	\begin{array}{cccc}
	1 & 0 & 0 & 0 \\
	0 & 1 & 0 & 0 \\
	0 & 0 & c_{34} & s_{34} \\
	0 & 0 & -s_{34} & c_{34} \\
	\end{array}
	\right)
	\left(
	\begin{array}{cccc}
	1 & 0 & 0 & 0 \\
	0 & c_{24} & 0 & e^{i \delta_2} s_{24} \\
	0 & 0 & 1 & 0 \\
	0 & -e^{-i \delta_2} s_{24} & 0 & c_{24} \\
	\end{array}
	\right)
	\left(
	\begin{array}{cccc}
	c_{14} & 0 & 0 & e^{i \delta_1} s_{14} \\
	0 & 1 & 0 & 0 \\
	0 & 0 & 1 & 0 \\
	-e^{-i \delta_1} s_{14} & 0 & 0 & c_{14} \\
	\end{array}
	\right)
	\left(
	\begin{array}{c|c}
	V_{3\times 3} & 0  \\
	\hline
	0 & 1 \\
	\end{array}
	\right)\,
}
for the $4\times 4$ diagonalizing matrix in the basis where $Y^u$ is diagonal.
The CKM matrix $V$ is obtained by chopping the last row.
We use the familiar shorthand where $c_{ij}=\cos\theta_{ij}$ and $s_{ij}=\sin\theta_{ij}$.
The $3\times 3$ block is the standard parametrization
\eq{
\label{ckm:standard}
V_{3\times 3}=
	\left(
	\begin{array}{ccc}
	1 & 0 & 0 \\
	0 & c_{23} & s_{23} \\
	0 & -s_{23} & c_{23} \\
	\end{array}
	\right)
	\left(
	\begin{array}{ccc}
	c_{13} & 0 & e^{-i \delta } s_{13} \\
	0 & 1 & 0 \\
	-e^{i \delta } s_{13} & 0 & c_{13} \\
	\end{array}
	\right)
	\left(
	\begin{array}{ccc}
	c_{12} & s_{12} & 0 \\
	-s_{12} & c_{12} & 0 \\
	0 & 0 & 1 \\
	\end{array}
	\right)\,.
}
This parametrization is interesting because $|V_{iB}|$ have very simple forms:
\eq{
\label{ViB:angles}
(|V_{uB}|,|V_{cB}|,|V_{tB}|)=(e^{i \delta_1} s_{14}, e^{i \delta_2} c_{14} s_{24}, c_{14}c_{24}s_{34})
\,.
}

In this parametrization for $V$ we can count six angles and three phases.
The number of phases match the known number of CP violating phases.
If we add the four down quark masses, including $M_B$, and the three up quark masses, we obtain the 16 parameters that match the number of Lagrangian parameters in \eqref{N.para:VLQ}.

The blue points in Fig.\,\ref{fig:X} represent the possible $|X^{d}_{ij}|$, 
using this parametrization, fixing the $3\times 3$ block \eqref{ckm:standard} to the 
best-fit values \eqref{ckm:bf}, varying $\theta_{i4}$ in the hierarchical range
\eqali{
\label{thetai4:range}
\theta_{14}&\in [3.70\times 10^{-6},\,1.69\times 10^{-4}]\,,~
\cr
\theta_{24}&\in [3.60\times 10^{-5},\,1.44\times 10^{-3}]\,,~
\cr
\theta_{34}&\in [7.15\times 10^{-4},\, 0.0273]\,,
}
while the phases $\delta_1,\delta_2$ are allowed any value in the whole range $[0,2\pi)$.
For better visualization and efficiency of point generation, we use an uniform distribution in $\log(\theta_{i4})$ instead of in $\theta_{i4}$ themselves.
We can see that the scatter plot of the blue points, representing the general case of VLQs using the ranges \eqref{thetai4:range}, mimics well the NB-VLQ case.
This study leads us to conclude that for $|X^d_{ij}|$ the strong correlations appearing in Fig.\,\ref{fig:X} basically follow from the hierarchical structure of $|V_{iB}|$ which is also possible for one generic VLQ by choosing the parameters appropriately.
It differs from the case of mixing only with the third family where all $X^d_{ds}=X^d_{db}=X^d_{sb}=0$ and only $X^d_{bB}$ is nonzero in accordance to \eqref{Xd:3rd.fam}.

\subsection{CP odd invariants}

In the SM where the CKM matrix $V$ is $3\times 3$, there is only one physical phase which describes all CP violating phenomena in the SM.
This sole phase sets the value of all quartic CP odd invariants that can be constructed from $V_{ij}$ and all of them are equal to the so called Jarlskog invariant of $V$\,\cite{jarlskog}, except for a sign ambiguity.
If formulated in terms of mass matrices, there is only one invariant as well\,\cite{branco.gronau}.

With the addition of one VLQ of down type, the CKM matrix becomes $3\times 4$ and two more physical phases appear, as in the explicit parametrization \eqref{V:param:angles}.
Therefore, there should be more than one independent Jarlskog invariant in this case.
Let us define the quartic CP odd invariants
\eq{
J_{ijkl}\equiv\im[V_{ij}V^\dag_{jk}V_{kl}V^\dag_{li}]\,.
}
For a $3\times 4$ matrix, the indices run from $i,k=1,2,3$ and $j,l=1,\dots,4$.

The properties
\eq{
J_{kjil}=-J_{ijkl}\,,\quad
J_{ilkj}=-J_{ijkl}\,,
}
allows us to choose $i<k$ and $j<l$.
The one-sided unitarity $VV^\dag=\id_3$ allows us to eliminate the $l=4$ invariants because
\eq{
J_{ijk4}=-J_{ijk1}-J_{ijk2}-J_{ijk3}\,.
}
For example,
\eq{
\label{J:4:ex}
J_{1124}=-J_{1122}-J_{1123}\,.
}
We are left with the 9 cases
\eq{
\label{ijkl}
(ijkl)\in\{(1122),(1123),(1132),(1133),(1223),(1233),(2132),(2133),(2233)\}
\,,
}
as the \emph{linearly independent} ones.
Since there are only three physical phases in total, there should be more \emph{algebraic} relations\,\cite{trautner} among them.\footnote{%
Ref.\,\cite{kielanowski} considered the SM with a fourth chiral family and concluded that, using an explicit parametrization for $V$, the vanishing of the first three $J_{1122},J_{1123},J_{1132}$ guarantee the vanishing of the rest.
This result should be valid for our case as well.
}
In terms of CP odd invariants depending on the mass matrices, Ref.\,\cite{saavedra.branco} gives invariance conditions in terms of seven invariants.
One simple example of algebraic relation in the SM is that $J^2$ is CP even and can be written in terms of $|V_{ij}|^2$\,\cite{branco:book}.\,\footnote{See, e.g., the case of invariants in the 2HDM\,\cite{trautner} for more complicated relations.}

Since the mixing of up-type quarks with $B_L$ are small and hierarchical, we expect that all $|J_{ijkl}|$ with $(ijkl)$ in the set \eqref{ijkl} would be close to the SM value\,\cite{ckmfitter} 
\eq{
\label{ckmfitter:J}
10^5\,J_{\rm SM}=3.060^{+0.071}_{-0.079}\,,
}
where
\eq{
\label{J:bf}
J_{\rm SM}=J_{1223}=J_{uscb}\,
}
is the most used Jarlskog invariant in the SM.

\begin{figure}[h]
\includegraphics[scale=0.5]{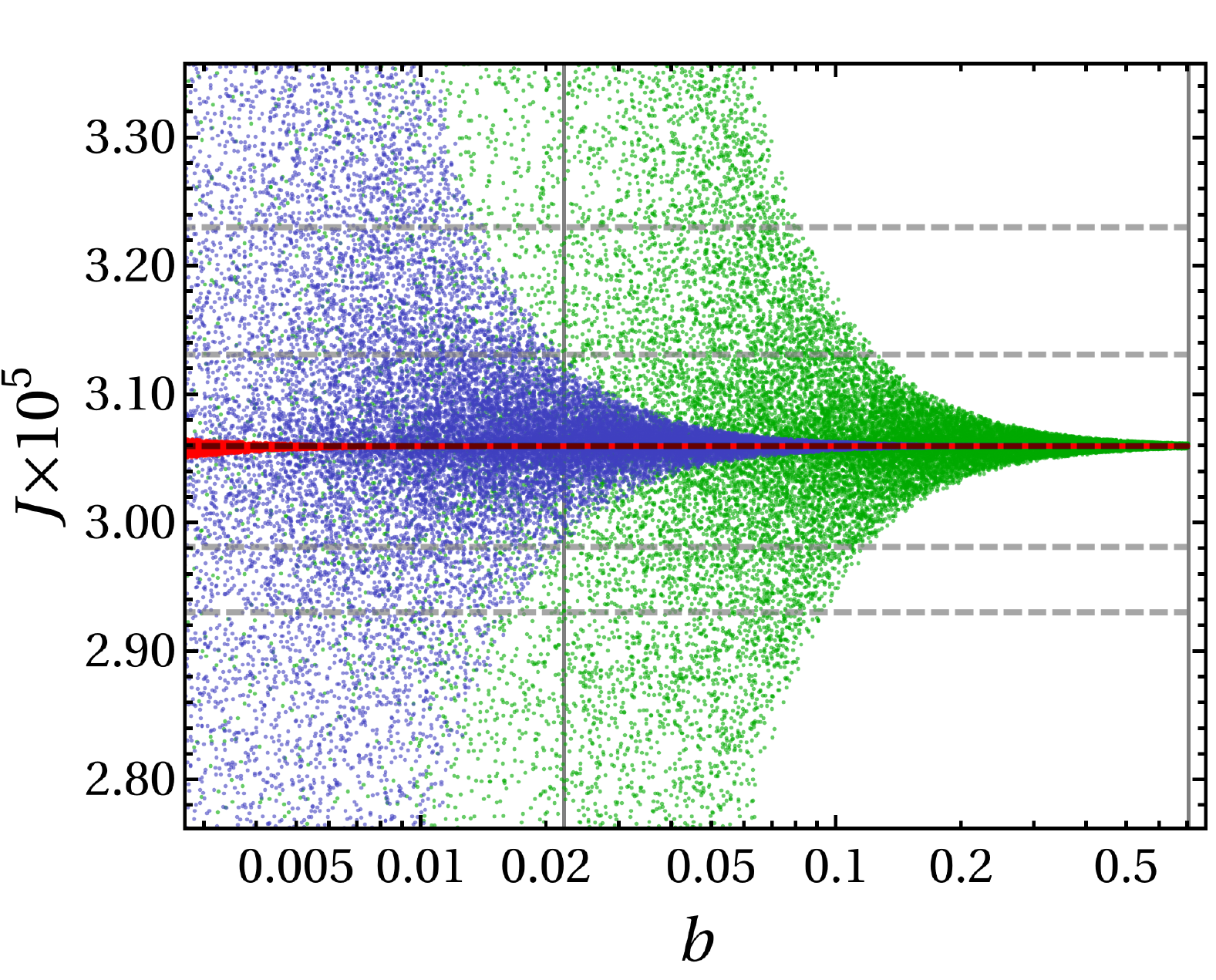}
\caption{\label{fig:J}
Scatter plot of $J_{2233}$ (green), $J_{1223}$ (blue) and $(-J_{1132})$ (red) as a function of $b$ for $M_B=1.3\,\unit{TeV}$.
The dashed lines mark the 1$\sigma$ and 2$\sigma$ intervals of CKMfitter\,\cite{ckmfitter}.
They enclose the black dashed line representing $J$ of the SM at the best-fit.
The vertical gray lines show the interval in \eqref{b:interval}.
The parameters $\beta_2,\beta_3,\gamma$ are varied through their whole range.
}
\end{figure}

Such an expectation is confirmed in Fig.\,\ref{fig:J} where we show three Jarlskog invariants in the set \eqref{ijkl}. 
They all scatter around the best-fit value in \eqref{J:bf} with increasing dispersion as $b$ decreases. We choose $J_{1223}$ and two other representatives, one with a very small dispersion ($J_{1132}$) and the other with the largest dispersion ($J_{2233}$).
The others have similar or intermediate behavior.
We also show in dashed gray lines the intervals of $1\sigma$ and $2\sigma$ for $J_{\rm SM}$ of CKMfitter\,\cite{ckmfitter}.
Also, the approximately equal values for all the $|J_{ijkl}|$ of \eqref{ijkl}
indicates that all Jarlskog invariants involving the index 4 are much smaller than the ones involving the $3\times 3$ block of the $V$.
In the example of eq.\,\eqref{J:4:ex}, the right-hand side would vanish in the SM and for one NB-VLQ the left-hand side shows a dispersion around zero.

The deviation of $J_{2233}$ from the SM value we see in Fig.\,\ref{fig:J} could be tested in the future by a more precise determination of $\phi_s$ proportional to the angle between $V_{cs}V_{cb}^*$ and $V_{ts}V_{tb^*}$ which enters precisely in $J_{2233}=\im[V_{cs}V^*_{ts}V_{tb}V^*_{cb}]$.
Currently, the errors are no smaller than 40\%\,\cite{phi_s} but the precision at LHCb at High-luminosity LHC with 300 fb$^{-1}$ is expected to be around 10\%.\,\cite{LHCb.high}.

One may note that there is no clear dependence (correlation) of the quartic CP odd invariants of Fig.\,\ref{fig:J} on the parameter $b$ which supposedly controls CP violation.
That happens because in \eqref{Yd:nB=1} we are fixing the $3\times 3$ block of the CKM matrix to match the SM one within the seesaw approximation.
Then matrix $\cY^d$ depends implicitly on $b$ through \eqref{cal.Yd:inverse}.
In other words, we are fixing the soft (spontaneous) CP violation of the model to mimic the explicit CP violation of the SM.
The limit $b\to 0$ does not lead to CP conservation.
If we were to make $\cY^d$ independent of $b$ and take the limit, CP conservation would be achieved and all CP odd invariants go to zero.

\section{Conclusions}
\label{sec:conclu}

We have defined and analyzed the SM augmented by vector-like quarks of Nelson-Barr type (NB-VLQs). These VLQs could be the lightest states beyond the SM arising from the solution to the strong CP problem through the Nelson-Barr mechanism in which CP is a fundamental symmetry only broken spontaneously.
Without access to the scalars that spontaneously break CP ---they may lie much above the scale of the VLQs--- this scenario can be defined by CP being softly broken by terms connecting the SM quarks with these new VLQs. In this scenario, this soft breaking is the origin of the CKM CP violation of the SM.

Due to the soft origin of CP breaking, models with NB-VLQs are described with one less parameter than a generic model with the same number of VLQs.
In special, for one NB-VLQ, only one CP odd quantity source all CP violation of the model compared to the total of three phases that appear for one generic VLQ without CP restriction.
Because the soft CP breaking needs to reproduce the explicit CP violation of the SM, 
the NB-VLQs cannot decouple completely and their coupling with the $Z$ through flavor changing currents cannot be made to vanish, although  they are allowed to be unobservably small.

For one NB-VLQ, we were able to solve the technical problem of parametrizing the model separating the ten parameters of the SM that can be chosen as input from the five new parameters describing the rest of the model, one of which is the new quark mass $M_B$.
This parametrization, denoted as the seesaw parametrization, assumes the leading quark seesaw approximation and is reliable as long as the quark seesaw is a good approximation.
Adopting $1\%$ as the maximum deviation allowed,
for $M_B=1.3\,\unit{TeV}$, we have found the lower limit of around 0.02 for the parameter $b$ that effectively controls the quality of the seesaw parametrization.
Concentrating on a VLQ of down type $B$, 
its mixing with the up-type quarks of the SM in the coupling with $W$ are hierarchical: the mixing is hierarchically larger the heavier is the SM quark. 
This feature can be seen in Fig.\,\ref{fig:ViB}.
Compared to the benchmark case of one VLQ, not of Nelson-Barr type, mixing only to the third family of the SM, significant deviations are possible.

Analyzing the possible flavor constraint for a NB-VLQ of down-type, the model is basically flavor safe due to the hierarchical mixing mentioned above.
The CP violation of the SM is also largely reproduced because the Jarlskog invariant cannot deviate much from the SM value.
The strongest constraint comes from $R_b$ followed by constraints from the kaon system.
This analysis can be seen in Sec.\,\ref{sec:flavor}.

In conclusion, models with NB-VLQs can be an interesting benchmark for a naturally flavor aligned VLQ model with one less free parameter than the generic version.
Strong correlations between various flavor observables appear and are possibly testable with enough precision.
For one NB-VLQ, an explicit parametrization was presented which can use to a good approximation the SM flavor parameters as input.
This will allow further detailed studies of this scenario.

\acknowledgments

The authors thank João Silva for helpful comments and Igor Ivanov for discussions about CP violating phases.
C.C.N.\ acknowledges partial support by Brazilian Fapesp, grant 2014/19164-6, and
CNPq, grant 304262/2019-6.
This study was financed in part by the Coordenação de Aperfeiçoamento de Pessoal de Nível Superior – Brasil (CAPES) – Finance Code 001.

\appendix
\section{Other formulas for partial diagonalization}
\label{ap:partial}

An expression for $W_R^{dd}$ alternative to \eqref{W:dd:1} can be obtained from the orthonormality of the leftmost column of blocks in \eqref{WR}, which yields
\eq{
\label{W:dd:2}
W_R^{dd}{W_R^{dd}}^\dag=\Big[\id_3-{\cM^{Bd}}^\dag (\cM^B\cM^{B\tp})^{-1}\cM^{Bd}\Big]^{-1}\,.
}
One can check the equivalence between \eqref{W:dd:1} and \eqref{W:dd:2} by computing $W_R^{dd}{W_R^{dd}}^\dag(W_R^{dd}{W_R^{dd}}^\dag)^{-1}=\id_3$.
If we choose $W_R^{dd}$ to be hermitian, we can write 
\eq{
\label{W:dd:herm}
W_R^{dd}=\Big(\id_3-{\cM^{Bd}}^\dag H_B^{-1}\cM^{Bd}\Big)^{1/2}\,.
}
Other choices are related by further unitary rotation from the right.

The block $W^{dB}_R$ in \eqref{WR} is $3\times n_B$ and can be parametrized with $n_B$ vectors $u_i$ as
\eq{
W^{dB}_R={\cM^{Bd}}^\dag {M^B}^{\dag -1}=\big(u_1\big|u_2\big|\dots\big|u_{n_B}\big)\,.
}
Being subparts of normalized vectors, they obey $|u_i|\le 1$. The limiting case $|u_k|=1$ means that $B_k$ decouples from the SM.
The parametrization \eqref{w} for $n_B=1$ is a special case.

Then $Y^B$ in \eqref{Yd:YB} can be written as
\eq{
Y^B=\cY^d\big(u_1\big|u_2\big|\dots\big|u_{n_B}\big)\,.
}
If we choose the basis where \eqref{W:dd:herm} is valid, we can also write
\eq{
Y^d=\cY^d\Big(\id_3-\sum_i u_iu_i^\dag\Big)^{1/2}\,.
}

\section{CP conserving limit}
\label{ap:contradiction}

Here we solve for $n_B=1$ the apparent contradiction coming from the presence of \emph{one} CP odd quantity \eqref{N.cp.odd:true} among the total of 15 parameters \eqref{N.param:NB} in the NB case compared to the 13 parameters \eqref{N.param:VLQ:real} in the CP conserved version.
One would expect that the difference $15-13=2$ would be the number of CP violating quantities.
We will see that such an expectation will not be realized due to the appearence of an additional reparametrization freedom.

Let us recall how the 15 parameters in the NB case are distributed in the basis where $\cM^{Bd} \sim (0,iy,x)$, cf.\,\eqref{w:a,b}: 
\eq{
\cY^u\sim 3,\quad 
\cY^d\sim 3+3+3\,,\quad \cM^{Bd}\sim 2\,,\quad \cM^B\sim 1.
}
We changed from $(a,b)$ to the easily related $(x,y)$ in $\cM^{Bd}$, where $x,y$ are real. The CP conserving limit is reached when $y\to 0$.
In this limit, not only we lose the CP violating parameter but we gain an $SO(2)$ freedom to rotate in the subspace $(d_{1R},d_{2R})$ which leaves $\cM^{Bd}$ invariant but allows us to \emph{remove} one parameter in $\cY^d$.
We end up with
\eq{
\cY^u\sim 3,\quad 
\cY^d\sim 3+3+3-1\,,\quad \cM^{Bd}\sim 1\,,\quad \cM^B\sim 1,
}
which matches 13.
So in the CP conserving limit of a theory with one NB-VLQ \textit{one CP even parameter} becomes \emph{unphysical}.
This is akin to the case of the SM where sending, e.g., $\theta_{23}\to 0$ in the CKM matrix, makes the CP phase $\delta$ becomes unphysical, i.e.,  it can be rephased away.
The freedom to remove one CP even parameter remains for $n_B>1$.

This example shows that the technique employed in Sec.\,\ref{sec:CPV} for counting the number of CP odd quantities (for the generic VLQ case) from the difference between the total number of parameters and the CP conserving limit must be accompanied with checks that discard the possibility of appearance of an additional reparametrization freedom.

\section{Minimization}
\label{ap:min}

Here we minimize the righthand side of \eqref{flavor.vio} to find the amount of irreducible flavor violation in NB-VLQ models.

For the norm, we use the Frobenius norm for a square matrix $A$,
\eq{
\norm{A}\equiv \sqrt{\Tr[A^\dag A]}\,.
}
We vary 6 parameters in $\cY^d$ in the parametrization
\eq{
\cY^d=O_{d_L}\hat{\cY}^d\,;
}
$O_{d_L}$ is a real orthogonal matrix (three mixing angles) and $\hat{\cY}^d$ is a diagonal matrix with three non-negative entries.
An orthogonal matrix on the right of $\hat{\cY}^d$ is irrelevant in this context.
We also vary the two phases in \eqref{majora-like} and thus we minimize the norm of the righthand side of \eqref{flavor.vio} with respect to the total number of 8 parameters.

The result is\,\footnote{%
If we had minimized only the off-diagonal part, $|A_{12}|^2+|A_{13}|^2+|A_{23}|^2$, with $A$ being the righthand side of \eqref{flavor.vio}, we would have obtained smaller (at most $3\times 10^{-12}$ in modulus) off-diagonal entries but larger (at least $10^{-5}$) entries in the diagonal.
}
\eqali{
\label{flavor.vio:min}
\Big(V_{d_L}^\dag\cY^d{\cY^d}^\tp V_{d_L}-(\hY^d)^2\Big)_{\min}
=
10^{-7}\times
\mtrx{0.851~~ & 1.298\, e^{- 0.88 i\pi} & 0.034\,e^{0.11 i\pi} \cr
* & 3.591 &  0.004\,e^{- 0.01 i\pi} \cr
* & * & 5.498}\,,
}
for
$\beta_2=1.903503$ and $\beta_3=1.903491$ in \eqref{majora-like} and 
\eq{
\label{cal-Yd:min}
\cY^d=\mtrx{0.9969 & 0.0787 & -0.0036 \cr
0.0788 &  -0.9960 &  0.0421 \cr
0.0002 & 0.0423 & 0.9991}\mtrx{2.403\times 10^{-4} &&\cr &8.279\times 10^{-4}&\cr &&2.404\times 10^{-2}}\,.
}
A possible orthogonal matrix on the right of \eqref{cal-Yd:min} is not determined by this procedure.

The matrix \eqref{flavor.vio:min} is positive definite and represents the minimal flavor violating matrix for $n_B\ge 3$, using the procedure above.
We have used $\hY^d=\diag(2.702\times 10^{-5},5.461\times 10^{-4},2.403\times 10^{-2})$ and the CKM matrix in the standard parametrization from Ref.\,\cite{PDG} for $\ckmsm$.

For a single NB-VLQ ($n_B=1$) of mass $M_B$, the relation between \eqref{flavor.vio} and $\delta X^d$ is direct:
\eq{
\delta X^{d} = \frac{v^{2}}{2 m_{B}^{2}}\left[V_{d_L}^\dag\cY^d{\cY^d}^\tp V_{d_L}-(\hY^d)^2\,\right].
}
Unfortunately, if we use the minimal values in \eqref{flavor.vio:min}, and take $M_B=1\,\unit{TeV}$, we obtain values no larger than roughly $10^{-9}$ and this level of flavor changing effects is far from detectable\,\cite{saavedra:flavor}.
Strictly speaking, the minimal values in \eqref{flavor.vio:min} are not valid for $n_B=1$, but the minimization procedure ensures that larger values would result for $n_B=1$ because additional constraints would be required.
If we naively translate this result to a model of a single up-type NB-VLQ, we would only gain two orders of magnitude due to the larger values of up-type SM yukawas.

\section{$\mu<1$}
\label{ap:mu}

Define the hermitian and positive definite matrix $H\equiv Y^d{Y^d}^{\dag}$ and separate it into its real and imaginary part:
\eq{
H=H_1+iH_2\,.
}
Then the expression in \eqref{cano:Im}, together with \eqref{def:mu}, can be rewritten as
\eq{
H_2=H_1^{1/2}\cO\,\mu\mtrx{0 &&\cr &0&-1\cr &1&0}\cO^\tp H_1^{1/2}\,.
}
Plugging this into $H$, we find
\eq{
H=H_1^{1/2}\cO\left[\id+i\mu\mtrx{0 &&\cr &0&-1\cr &1&0}\right]\cO^\tp H_1^{1/2}\,.
}
The positive definiteness of $H$ implies that the inner matrix inside brackets should be positive definite, so $1-\mu^2>0$.


\end{document}